\documentclass[usenatbib]{mnras}

\usepackage{mathptmx}            
\usepackage{ae,aecompl}          
\usepackage{amssymb}             
\usepackage[T1]{fontenc}         
\usepackage{graphicx}            
\usepackage{booktabs}            

\makeatletter
\def\ScaleWidthIfNeeded{%
 \ifdim\Gin@nat@width>\linewidth
    \linewidth
  \else
    \Gin@nat@width
  \fi
}
\makeatother
\setkeys{Gin}{width=\ScaleWidthIfNeeded,keepaspectratio}%

\providecommand{\tightlist}{%
  \setlength{\itemsep}{0pt}\setlength{\parskip}{0pt}}

\usepackage[utf8]{inputenc}

\input{binhex}
\makeatletter
\def\uc@dclc#1#2#3{%
  \ifnum\pdfstrcmp{#2}{mathletters}=\z@
    \begingroup\edef\x{\endgroup
      \noexpand\DeclareUnicodeCharacter{\hex{#1}}}\x{#3}%
  \fi
}
\input{uni-3.def}
\def\uc@dclc#1#2#3{%
  \ifnum\pdfstrcmp{#2}{default}=\z@
    \begingroup\edef\x{\endgroup
      \noexpand\DeclareUnicodeCharacter{\hex{#1}}}\x{#3}%
  \fi
}
\input{uni-34.def}
\makeatother

\DeclareUnicodeCharacter{00D7}{\times}

\title[
Thermal effects in super-Earth interiors
]{
In hot water: effects of temperature-dependent interiors on the radii of
water-rich super-Earths
}

\author[
Thomas \& Madhusudhan
]{
Scott W. Thomas \&
Nikku Madhusudhan\\
Institute of Astronomy, University of Cambridge, Madingley Road,
Cambridge CB3 0HA, United Kingdom
}

\date{January 2016}
\pubyear{2016}

\begin{document}
\label{firstpage}
\pagerange{\pageref{firstpage}--\pageref{lastpage}}
\maketitle

\begin{abstract}
    Observational advancements are leading to increasingly precise
    measurements of super-Earth masses and radii. Such measurements are used
    in internal structure models to constrain interior compositions of
    super-Earths. It is now critically important to quantify the effect of
    various model assumptions on the predicted radii. In particular, models
    often neglect thermal effects, a choice justified by noting that the
    thermal expansion of a solid Earth-like planet is small. However, the
    thermal effects for water-rich interiors may be significant. We have
    systematically explored the extent to which thermal effects can
    influence the radii of water-rich super-Earths over a wide range of
    masses, surface temperatures, surface pressures and water mass
    fractions. We developed temperature-dependent internal structure models
    of water-rich super-Earths that include a comprehensive
    temperature-dependent water equation of state. We found that thermal
    effects induce significant changes in their radii. For example, for
    super-Earths with 10 per cent water by mass, the radius increases by up
    to 0.5\(\,\)R\(_⊕\) when the surface temperature is increased from 300
    to 1000\(\,\)K, assuming a surface pressure of 100\(\,\)bar and an
    adiabatic temperature gradient in the water layer. The increase is even
    larger at lower surface pressures and/or higher surface temperatures,
    while changing the water fraction makes only a marginal difference.
    These effects are comparable to current super-Earth radial measurement
    errors, which can be better than 0.1\(\,\)R\(_⊕\). It is therefore
    important to ensure that the thermal behaviour of water is taken into
    account when interpreting super-Earth radii using internal structure
    models.
\end{abstract}

\section{Introduction}\label{introduction}

One of the most interesting classes of planets today is the class of
super-Earths, planets with masses between 1 and 10\(\,\)M\(_⊕\). With no
analogues in the solar system, it is not known whether they are scaled
up rocky planets or scaled down Neptunes. About 40 super-Earths with
measured masses and radii are currently known. Their radii range from 1
to 7\(\,\)R\(_⊕\).\footnote{This number is taken from the
  \href{http://www.exoplanets.org}{exoplanets.org} database (confirmed
  planets only).} With the potential to have moderate atmospheres and
plate tectonics, super-Earths represent an important class of planets in
the broader context of planetary diversity and planetary habitability
\citep{Haghighipour2011, Baraffe2014}.

Recent observational advancements are leading to increasingly precise
measurements of masses and radii of these small planets. Such
measurements are being used with internal structure models to place
constraints on the interior compositions of super-Earths. Many planets
are well-described by multi-layer models consisting of iron, silicates,
and water
\citep[e.g.][]{Valencia2006, Fortney2007, Sotin2007, Seager2007} and
others have included layers of hydrogen or other volatiles to explain
the inflated radii of some super-Earths
\citep[e.g.][]{Rogers2010, Lopez2012}. Given the high-precision radii
measurements, it is now critically important to quantify the dependence
of predicted radii of super-Earth models on the various model
assumptions.

Our goal is to quantify the effects of temperature-dependent internal
structure models on the predicted radii of super-Earths. An
understanding of the effects of temperature on the internal structures
of planets is especially relevant as our observational capabilities for
measuring radii and masses improve. In particular, we are interested in
understanding to what degree the observable radius of a planet may be
affected by thermal expansion of its interior. We use water as a case
study for this, focusing on super-Earth planets consisting of a rocky
Earth-like core underneath a heated water layer.

The remainder of this section provides an overview of how these
planetary interior models can be useful, why we expect
temperature-dependent models to be different and why water-rich planets
make interesting test cases for assessing whether this temperature
dependence is significant.

\subsection{Planetary interior models}\label{planetary-interior-models}

As atmospheric characterisation techniques improve, the question of what
lies beneath the atmosphere has naturally arisen. We care about
planetary interiors because they are linked to the formation history of
the planet, because they are shaped by and shape the planetary
atmosphere and because they are key to answering questions about
habitability \citep{Sotin2010}. Understanding these exoplanets also
allows us to place our own Earth into context: how unique are we? We
therefore seek to understand, if not the interiors of individual
exoplanets, at least something about broad classes of planets. But it is
here that we are confronted by a lack of data, because we have very
little ability to directly probe the interiors of exoplanets.

This lack of a rich source of observational data for planetary interiors
means that we rely strongly on models. Even inside our solar system, our
knowledge of planetary interiors is limited by the indirect ways in
which we can probe them. On Earth we have the advantage of seismic
measurements, and in our solar system we have various gravitational
moments to constrain interior structures. Outside the solar system we
have only the masses and radii of planets to work with. Models from
first principles (numerical or analytical models based on the physics of
solid and liquid spheres) therefore dominate the field.

Planetary interior models are a worthwhile starting point to make sense
of the limited observational data we have. These models are inspired by
earlier successes with stellar structure models, which are key to
interpreting observations of stars. Others had previously considered the
internal structures of planets in our solar system \citep[for
example,][]{Hubbard1980}, but the study often taken as the base for
planetary interior modelling is by \citet{Zapolsky1969} who constructed
mass--radius relations for large homogeneous isothermal spheres. A
number of internal structure models have been developed for exoplanets,
starting with early works a decade ago
\citep[e.g.][]{Valencia2006, Fortney2007, Sotin2007, Seager2007}. The
ever-increasing number of known exoplanets, many of which have both mass
and radius measurements, are a diverse and interesting set of objects to
which to apply these models.

The first way in which planetary interior models can be useful is to
make broad inferences about the structure of a planet. There is some
information available about any planet despite an inherent degeneracy
between different compositions. We can immediately exclude certain
classes of models: for example, small planets with large radii must
almost certainly have large hydrogen envelopes. We can also take more
sophisticated approaches. \citet{Sotin2007} modelled planets by fixing
their compositions based on the properties of the host star.
\citet{Madhusudhan2012a} argued for a carbon-rich interior in the
exoplanet 55 Cnc e based on its carbon abundance and on its density
matching that of pure carbon. \citet{Dorn2015} also showed that mass and
radius alone can constrain the size of a planet's core if we assume it
is pure iron.

We can also hope to make progress in a statistical sense by examining
populations of planets. Such progress is possible even if we are unable
to pin down the precise structure of an individual planet. There are
promising advances in this direction already. These usually involve
inverse Bayesian analyses. For example, \citet{Rogers2014} investigated
the size demographics of planetary populations and set an approximate
boundary of 1.6\(\,\)R\(_⊕\) beyond which planets are likely to have
gaseous envelopes.

Finally, interior structure models may be useful when combined with
prescriptions for planetary formation. \citet{Mordasini2012} took this
approach, combining interior structure calculations with models of the
protoplanetary disk to produce synthetic populations of planets.
\citet{Lopez2012} also made model planets and explored how they evolve
and lose mass through time \citep[see also][]{Owen2015}.

If we are to use mass and radius to constrain the interior structure of
a planet, we should ensure that our models are precise and accurate. But
more importantly, we should understand where our models need to be
precise and accurate and where such effort is wasted. We therefore
require a thorough understanding of what factors can affect the
mass--radius relation. We also need to know to what extent we are able
to invert the relation to determine a composition.

The internal structure of a planet is not well-constrained by its mass
and radius alone \citep{Rogers2010}. However, we know that we can obtain
some compositional constraints from observations of the planet and its
host star. Above, we mentioned works by \citet{Sotin2007} and
\citet{Madhusudhan2012a}, who used host star information in this way.
\citet{Dorn2015} also used probabilistic models, incorporating the host
star chemical abundances, to conclude that ``uncertainties on mass,
radius, and stellar abundance constraints appear to be equally
important.'' \citet{Grasset2009} indicated the need for good radius
measurements, especially for dry silicate-rich planets for which
numerical models can provide radius estimates to precisions of less than
5 per cent. \citet{Unterborn2015} used a mineral physics toolkit to
perform a sensitivity analysis for rocky super-Earths, concluding that
the mass--radius relationship is most strongly altered by the core
radius and the presence of light elements in the core.

The presence of an atmosphere could also contribute significantly to the
observed radius. \citet{Rogers2010a} have modelled isothermal
super-Earth interiors overlaid by a volatile atmosphere. Additionally,
\citet{Valencia2013} considered coupled atmosphere--interior models,
which also included atmospheric mass loss, and explored the dependence
of radii on various model parameters such as the irradiation, water
content and metallicity. The effect of an atmosphere is important,
especially given that observations can probe spectral ranges where
atmospheric absorption could be significant \citep{Madhusudhan2015}.

Though the factors above are all important, the effect of temperature on
the mass--radius relation has not been thoroughly explored \citep[but
see e.g.][]{Valencia2013}. This is for several reasons. First, its
effects are thought to be relatively minor in the first place:
\citet{Howe2014} estimate that the effect of thermal corrections on an
iron-silicate planet's radius is approximately 5 per cent.
\citet{Grasset2009} also describe how the radius of an Earth-like planet
is not strongly affected by temperature changes. If the effect is small
compared with current observational uncertainties, it is not necessarily
relevant. Secondly, modelling is easier if we assume zero-temperature or
isothermal spheres of material, because we do not have to deal with
energy transport within the planet. Finally, the data on thermal
expansion of heavy elements are sparse at the high temperatures and
pressures characteristic of planetary interiors \citep{Baraffe2008}.
Therefore mass--radius relations or models of individual planets
traditionally had no temperature dependence at all
\citep{Zapolsky1969, Seager2007} but it is increasingly being included
and thermal effects on radii are being explored \citep[for example,
see][]{Valencia2013}.

\subsection{Temperature dependence of water-rich
planets}\label{temperature-dependence-of-water-rich-planets}

The degree to which thermal structure may affect the properties of a
water-rich planet has not yet been well studied. Super-Earth planets
with significant water layers, sometimes called waterworlds, provide an
interesting testbed for our investigation. They may display more
significant variation in their properties, both observable and internal,
than purely Earth-like (iron and silicate) planets. They are therefore a
worthwhile target for this study.

\begin{figure}
\centering
\includegraphics{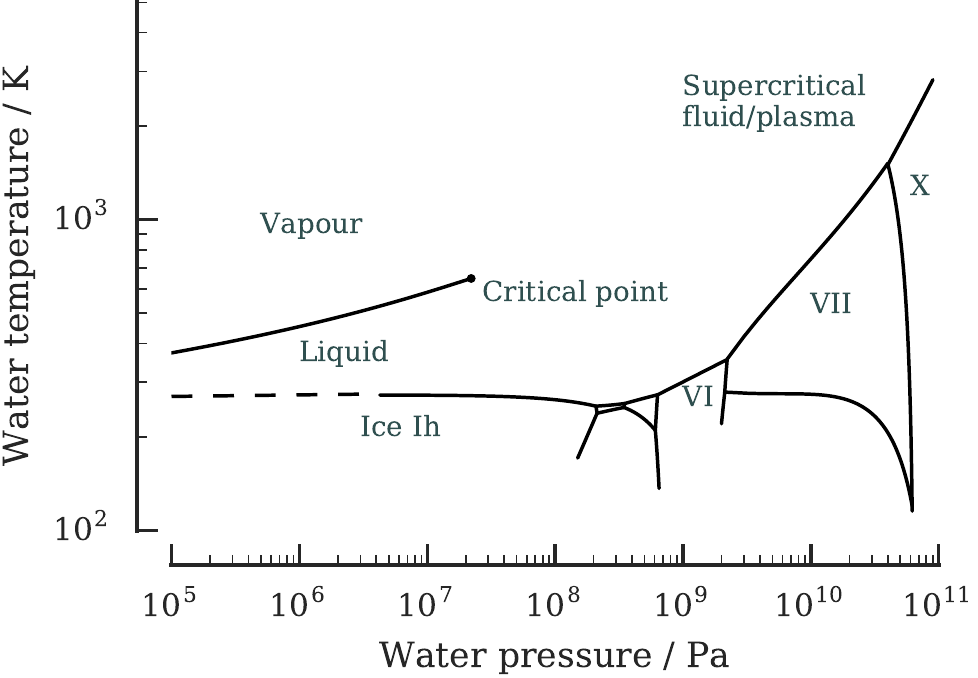}
\caption{Phase diagram of water. Water has a rich and interesting phase
structure. Here we show some of the key phases which are relevant when
modelling a watery planet: liquid, vapour, and solid ice Ih, but also
more exotic phases such as the high-pressure ices. Lines mark the
boundaries of each phase as given by \citet{Choukroun2007} and
\citet{Wagner2002}.\label{fig:water-phases}}
\end{figure}

Water presents an opportunity to assess thermal effects in a material
that has a rich and interesting phase structure across a large
temperature and pressure range (Fig. \ref{fig:water-phases}). At low
temperature and pressure, water exists as a liquid, vapour, or solid
(Ice Ih). At high pressure, it takes on a number of alternate ice forms
(Ice V, VI, VII, X, etc.) \citep{Choukroun2007}. It can also exist as a
low-density supercritical fluid or superheated vapour. This all means
that the behaviour of water layers is thermally interesting. The
behaviour of water is also strongly linked to questions of habitability
because Earth-sized solid planets with oceans provide the best
approximation to the one planet known to harbour life.

Others have previously investigated the structures of planets containing
a significant water component. For example, \citet{Ehrenreich2006}
studied the internal structure of the exoplanet OGLE 2005-BLG-390Lb,
modelling the phase changes throughout. \citet{Zeng2014} chose to
explore evolutionary effects, following the phase transitions within
model water-rich planets. They comment that ``{[}phase{]}
transformations may have a significant effect on the interior convective
pattern and also the magnetic field of such a planet, but they may only
affect the overall radius slightly.'' Our present study addresses the
question of exactly how much temperature variations affect the structure
and radius of water-rich planets and whether such effects are
observable.

\section{Method}\label{method}

Guided by the motivations above, we quantified the thermal effects in
super-Earths which contain significant amounts of water. By
\emph{thermal effects} we mean three effects in particular. First, we
mean the thermal expansion of a heated water layer on the surface or
within the planet's interior: this contrasts with models which treat the
planets as cold spheres. Secondly, we mean any temperature gradient
established within the planet: this is in contrast to the isothermal
case. And thirdly, we expect phase transitions within the water layer if
the temperature and pressure cross one of the boundaries between
different phases of water seen in Fig. \ref{fig:water-phases}.

To quantify these effects required several steps. First, we selected an
appropriate temperature-dependent equation of state. We built planetary
interior structure models that included this equation of state. We
incorporated a realistic temperature gradient into these models.
Finally, we explored the model parameter space. In particular, we
compared the mass--radius relationships for these water worlds across a
range of surface pressures, surface temperatures and interior
compositions.

In this section, we first explain how we built models of super-Earth
interiors: we constructed layered one-dimensional models of water, rock
and iron in varying proportions. We discuss our approach to the
temperature structure: we treated the temperature gradient as adiabatic
and self-consistently calculated it from the equation of state, an
approach which naturally handles phase boundaries within the water
layer. We explain how we used these models to construct
temperature-dependent mass--radius relations for both homogeneous and
layered planets. Next we present the equation of state for water that we
used. It is comprehensive over the pressure and temperature range
relevant to super-Earth interiors. We highlight the difficulty of
dealing with disparate sources of experimental and theoretical data in
different phases, as well as sources of uncertainty within the equation
of state. Finally, we present comparisons with previous works to verify
that our structural modelling code works appropriately.

\subsection{Interior structure
modelling}\label{interior-structure-modelling}

We constructed temperature-dependent planetary interior models of
water-rich super-Earths. We considered the planet to be spherically
symmetric, non-rotating and non-magnetic. The following equations govern
the structure of such a planetary interior.

\subsubsection{Planetary structure
equations}\label{planetary-structure-equations}

The mass continuity equation,
\begin{equation} {dr \over dm} = {1 \over 4πr^2ρ}, \label{eq:mass-continuity}\end{equation}
links \(r\), the radius of a spherical shell, to the mass \(m\) interior
to the shell and the density \(ρ\) of the shell. The equation of
hydrostatic equilibrium,
\begin{equation} {dP \over dm} = -{Gm \over 4πr^4}, \label{eq:pressure-gravity}\end{equation}
where \(P\) is the pressure at the shell and \(G\) is the gravitational
constant, ensures a balance of pressure and gravity. The equation of
state, \begin{equation} ρ = ρ(P, T), \label{eq:eos}\end{equation} is
used to calculate the density of the material in question from its
pressure and temperature \(T\).

Previous models of super-Earth interiors have treated temperature
gradients within the planet in a number of ways. From simple to complex,
these include:

\begin{enumerate}
\def\labelenumi{\arabic{enumi}.}
\tightlist
\item
  Isothermal models, which use only the equations above and an
  isothermal equation of state of the form \(ρ = ρ(P)\). This is the
  approach taken by \citet{Seager2007}.
\item
  Simple temperature prescriptions:

  \begin{enumerate}
  \def\labelenumii{\alph{enumii})}
  \tightlist
  \item
    One may assume a fixed temperature--pressure relation \(T = T(P)\)
    so as to reduce the equation of state to the form \(ρ = ρ(P)\) and
    then use the equations above. For example, \citet{Zeng2013} chose
    the melting curve of water for this purpose.
  \item
    Or one may choose a temperature profile \(T = T(r)\) for the planet
    (perhaps scaled appropriately to an internal or external boundary
    temperature) and then use the equations above.
  \end{enumerate}
\item
  An adiabatic or conductive temperature gradient or some combination of
  the two. For example, \citet{Valencia2010} used a convective interior
  with conductive boundary layers.
\item
  A full treatment, which adds an energy transport equation to the
  equations above then self-consistently solves this with a prescription
  for luminosity. For example, \citet{Wagner2011} modelled an adiabatic
  core underneath a radiogenically heated mantle.
\end{enumerate}

We did not explicitly handle energy transport in the manner of the
fourth option. Instead we chose the third approach and assumed an
adiabatic (isentropic) temperature gradient throughout the planet. The
equation for the adiabatic temperature gradient, as given by
\citet{Milone2014}, is
\begin{equation} { dT \over dr} = -{T α g \over c_p}. \label{eq:temp-gradient-radial}\end{equation}
where \(g = Gm/r^2\) is the gravity at the shell, \(c_p\) is the
isobaric heat capacity and \(α\) is the volumetric thermal expansion
coefficient. This is sometimes written \(α_V\) and is defined as the
fractional increase in volume per unit temperature increase,
\begin{equation} α = \left. {1 \over V}{∂V \over ∂T} \right|_p
     = - \left. {∂ \ln ρ \over ∂ \ln T} \right|_p. \label{eq:thermal-expansion}\end{equation}
Our sources for these latter two coefficients are detailed in our
equation of state section. Equation \ref{eq:temp-gradient-radial}
combined with equation \ref{eq:mass-continuity} gives the temperature
gradient in terms of the mass co-ordinate,
\begin{equation} {dT \over dm} = -{T α G m \over ρ c_p 4 π r^4}. \label{eq:temp-gradient}\end{equation}

\subsubsection{Solving the structural
equations}\label{solving-the-structural-equations}

Together, equations \ref{eq:mass-continuity}, \ref{eq:pressure-gravity},
\ref{eq:eos} and \ref{eq:temp-gradient} define a structural model: three
ordinary differential equations and an equation of state linking
pressure, temperature and density. The choice of how to solve this
system depends on one's aim. A common approach has been to treat it as a
boundary value problem; that is, to integrate the structural equations
from initial conditions at the surface or centre of the planet. For
example, \citet{Seager2007} approached the isothermal problem (equations
\ref{eq:mass-continuity}, \ref{eq:pressure-gravity} and \ref{eq:eos}
only) from the inside out, choosing appropriate central pressures at the
\((r=0, m=0)\) boundary and building their models outward from there. We
instead integrate from the outside in, an approach taken by several
others \citep[e.g.][]{Rogers2010, Madhusudhan2012a}. This has the
advantage of allowing us to specify the surface temperature and pressure
as boundary conditions. These surface boundary conditions are more
closely linked to observable parameters than the central pressure and
temperature.

We used a Lagrangian system, where the mass interior to a given shell is
the independent variable; this is reflected in equations
\ref{eq:mass-continuity}, \ref{eq:pressure-gravity} and \ref{eq:eos}. It
is in contrast to the Eulerian co-ordinate system used by
\citet{Seager2007}, who take the radius \(r\) as the independent
variable. \citet{Rogers2012} claim that this formulation is more stable
under numerical integration. We also found it more convenient to be able
to specify differentiated planets in terms of mass fractions rather than
radial distances.

We solved this boundary value problem using a shooting method\footnote{For
  an example implementation, see \citet{Press2007}, chapter 18.}. This
method used a series of trial solutions, adjusting the initial
conditions as necessary based on the difference between the expected and
actual values at the end of the integration domain. For the initial
trial solution, we specified the surface boundary conditions: total
planetary mass \(M\), surface pressure \(P(M)\), and surface temperature
\(T(M)\). We also specified a search bracket for the radius
\([R_1(M), R_2(M)]\). Our code\footnote{Our code was written in
  \href{http://www.julialang.org}{Julia}.} used a fixed-step
fourth-order Runge--Kutta integrator to solve the system of differential
equations above. In each successive trial, it iteratively adjusted the
radius boundary condition \(R(M)\) according to the bisection
root-finding method to ensure that the radius approached zero as the
mass approached zero. We further required that \(r\) remained positive:
this avoided any numerical difficulty arising from the behaviour of the
equations at \(r=0\). We deemed the system to be converged acceptably
when the central radius \(r(m=0)\) was between 0 and 100\(\,\)m.

\subsubsection{Mass--radius relations with thermal expansion and
multiple
layers}\label{massradius-relations-with-thermal-expansion-and-multiple-layers}

We used our models to produce mass--radius relations for homogeneous
spheres of water as well as differentiated multi-layer models. We did
this first for the homogeneous isothermal case \citep[in the vein
of][]{Zapolsky1969} and then extended our models to include an adiabatic
temperature gradient. Our differentiated multi-layer models included a
water layer on top of a silicate mantle and an iron core. To do this,
they treated the equation of state, equation \ref{eq:eos}, as piecewise
in the mass co-ordinate. For example, consider a model which has a 5 per
cent (by mass) water layer on top of a silicate mantle. For this model,
we begin by evaluating equation \ref{eq:eos} using the water equation of
state. We then switch to the silicate equation of state once \(m\), the
mass interior to the spherical shell in equations
\ref{eq:mass-continuity}, \ref{eq:pressure-gravity} and
\ref{eq:temp-gradient}, drops below 95 per cent of the planetary mass.
It is possible to choose the integration grid such that this occurs
exactly at the end of an integration step. However, in practice a
sufficiently fine grid is also acceptable.

We ignored thermal effects within the iron and silicate layers. The
effect of thermal expansion in these solids is thought to be low
\citep[see][]{Seager2007, Grasset2009}. Including the expansion effects
of these materials would be very simple, but we have not yet collated
the equation of state data which would enable us to do so. Because we
ignored thermal expansion in these layers, we modelled them as
isothermal, which follows from setting \(α = 0\) in equation
\ref{eq:temp-gradient} so that \(dT / dm = 0\).

We aimed to accurately capture the density change of water at its phase
boundaries. Our equation of state for water therefore included its phase
transitions, which appear as density discontinuities in
pressure--temperature space. When calculating the adiabatic temperature
profile, we enforced temperature and pressure continuity at these phase
boundaries. We did this by ensuring that the equation for the adiabatic
temperature gradient, equation \ref{eq:temp-gradient}, was finite and
continuous. This effectively split the adiabatic temperature profile
into several different sections, consisting of one separate adiabat for
each phase and meeting at the phase boundaries of water. By handling
each phase separately, we avoided the numerical difficulty of taking a
derivative (equation \ref{eq:thermal-expansion}) across a density
discontinuity.

We note that it is possible to fix the radius and let another parameter
vary instead. This could be the mass, surface temperature, surface
pressure or the position of a layer boundary within the planet. Other
studies have used this approach to infer potential compositions for
planets of known mass and radius. We instead left the radius free to
investigate how much it was affected by the water layer's thermal
expansion. This was the primary goal of this study.

\subsection{Equation of state}\label{equation-of-state}

As the goal of this study was to investigate thermal effects, we
required a temperature-dependent equation of state for water. This
allowed us to treat thermal expansion self-consistently in our models.
We synthesized an equation of state for water from the best available
experimental and theoretical data over a wide range of pressure and
temperature.

\subsubsection{Previous equations of state for
water}\label{previous-equations-of-state-for-water}

There was no single comprehensive equation of state data set available
for water over the entire pressure and temperature range relevant to
super-Earth interiors. Previous studies have approached this problem by
stitching together equations of state which are valid for different
pressure regimes. For example, \citet{Seager2007} took this approach
with water, combining three temperature-independent equations of state
for ice VII:

\begin{itemize}
\tightlist
\item
  the Birch--Murnaghan equation of state at low pressures,
\item
  density functional theory calculations at intermediate pressures and
\item
  the Thomas--Fermi--Dirac model at very high pressures.
\end{itemize}

The piecewise function defined in this way is appropriate across a wide
pressure range.

This pressure piecewise approach neglects temperature dependence in the
equation of state but provides a robust approximation that is easy to
evaluate. In some cases, stitching the data in this fashion has revealed
that a simpler functional form works just as well. For example, this is
the case in the ``polytropic equation of state'' used by
\citet{Seager2007}. Such simple functional forms for the equation of
state have been used successfully to model planets as cold spheres since
the work of \citet{Zapolsky1969}. In other cases, a more detailed
functional form is needed to capture the behaviour of the material
fully; this is especially true if it undergoes phase transitions. For
example, the IAPWS formulation described by \citet{Wagner2002} uses a
complicated series of equations fitted to various sources of
experimental data for the behaviour of water in the vapour and liquid
phases.

In choosing equations of state, previous authors have taken similar
approaches to the stitching technique above. Although the choice of the
exact equations has varied as new experimental data were released, few
of these studies included thermal expansion. \citet{Howe2014} provided a
comprehensive overview of the equations of state chosen in previous
works to model planetary interiors. They included several different
materials of interest for planetary interiors: water ice, iron, and
silicates. We repeated this exercise, focusing exclusively on the water
equations of state across all its phases. Table \ref{tbl:eos-sources}
summarises our findings.\footnote{Where abbreviations are used in this
  table and Table \ref{tbl:our-eos}, Table \ref{tbl:eos-abbrevs}
  indicates from which studies they come.}

\begin{table}
\centering
\caption{Previous studies on planetary interior structures use a variety
of equations of state for water.\label{tbl:eos-sources}}
\begin{tabular}{@{}ll@{}}
\toprule
\begin{minipage}[b]{0.32\columnwidth}\raggedright\strut
Work(s)
\strut\end{minipage} &
\begin{minipage}[b]{0.62\columnwidth}\raggedright\strut
Water equation of state used
\strut\end{minipage}\tabularnewline

\midrule
\begin{minipage}[t]{0.32\columnwidth}\raggedright\strut
\citet{Baraffe2008}; \citet{Baraffe2014}
\strut\end{minipage} &
\begin{minipage}[t]{0.62\columnwidth}\raggedright\strut
TFD, BME, MGD
\strut\end{minipage}\tabularnewline
\begin{minipage}[t]{0.32\columnwidth}\raggedright\strut
\citet{Fortney2007}
\strut\end{minipage} &
\begin{minipage}[t]{0.62\columnwidth}\raggedright\strut
Simple power law from \citet{Hubbard1980}
\strut\end{minipage}\tabularnewline
\begin{minipage}[t]{0.32\columnwidth}\raggedright\strut
\citet{Fortney2009}
\strut\end{minipage} &
\begin{minipage}[t]{0.62\columnwidth}\raggedright\strut
H\(_2\)O-REOS
\strut\end{minipage}\tabularnewline
\begin{minipage}[t]{0.32\columnwidth}\raggedright\strut
\citet{Grasset2009}
\strut\end{minipage} &
\begin{minipage}[t]{0.62\columnwidth}\raggedright\strut
MGD; Vinet; BME; TFD; ANEOS; \citet{Belonoshko1991}
\strut\end{minipage}\tabularnewline
\begin{minipage}[t]{0.32\columnwidth}\raggedright\strut
\citet{Guillot1999}
\strut\end{minipage} &
\begin{minipage}[t]{0.62\columnwidth}\raggedright\strut
\citet{Hubbard1989}
\strut\end{minipage}\tabularnewline
\begin{minipage}[t]{0.32\columnwidth}\raggedright\strut
\citet{Howe2014}
\strut\end{minipage} &
\begin{minipage}[t]{0.62\columnwidth}\raggedright\strut
Vinet
\strut\end{minipage}\tabularnewline
\begin{minipage}[t]{0.32\columnwidth}\raggedright\strut
\citet{Hubbard1980}
\strut\end{minipage} &
\begin{minipage}[t]{0.62\columnwidth}\raggedright\strut
Simple power law
\strut\end{minipage}\tabularnewline
\begin{minipage}[t]{0.32\columnwidth}\raggedright\strut
\citet{Hubbard1989}
\strut\end{minipage} &
\begin{minipage}[t]{0.62\columnwidth}\raggedright\strut
Exponential polynomial EOS without temperature dependence
\strut\end{minipage}\tabularnewline
\begin{minipage}[t]{0.32\columnwidth}\raggedright\strut
\citet{Lopez2012}
\strut\end{minipage} &
\begin{minipage}[t]{0.62\columnwidth}\raggedright\strut
H\(_2\)O-REOS
\strut\end{minipage}\tabularnewline
\begin{minipage}[t]{0.32\columnwidth}\raggedright\strut
\citet{Madhusudhan2012}
\strut\end{minipage} &
\begin{minipage}[t]{0.62\columnwidth}\raggedright\strut
BME
\strut\end{minipage}\tabularnewline
\begin{minipage}[t]{0.32\columnwidth}\raggedright\strut
\citet{More1988}
\strut\end{minipage} &
\begin{minipage}[t]{0.62\columnwidth}\raggedright\strut
Quotidian EOS (ion EOS with Thomas--Fermi model)
\strut\end{minipage}\tabularnewline
\begin{minipage}[t]{0.32\columnwidth}\raggedright\strut
\citet{Nettelmann2008}
\strut\end{minipage} &
\begin{minipage}[t]{0.62\columnwidth}\raggedright\strut
LM-REOS
\strut\end{minipage}\tabularnewline
\begin{minipage}[t]{0.32\columnwidth}\raggedright\strut
\citet{Nettelmann2011}
\strut\end{minipage} &
\begin{minipage}[t]{0.62\columnwidth}\raggedright\strut
H\(_2\)O-REOS
\strut\end{minipage}\tabularnewline
\begin{minipage}[t]{0.32\columnwidth}\raggedright\strut
\citet{Redmer2011}
\strut\end{minipage} &
\begin{minipage}[t]{0.62\columnwidth}\raggedright\strut
\citet{French2009}
\strut\end{minipage}\tabularnewline
\begin{minipage}[t]{0.32\columnwidth}\raggedright\strut
\citet{Rogers2010a}
\strut\end{minipage} &
\begin{minipage}[t]{0.62\columnwidth}\raggedright\strut
IAPWS; IAPWS extrapolations; TFD
\strut\end{minipage}\tabularnewline
\begin{minipage}[t]{0.32\columnwidth}\raggedright\strut
\citet{Seager2007}; \citet{Rogers2010}; \citet{Zeng2014}
\strut\end{minipage} &
\begin{minipage}[t]{0.62\columnwidth}\raggedright\strut
Low-temperature polytropic EOS
\strut\end{minipage}\tabularnewline
\begin{minipage}[t]{0.32\columnwidth}\raggedright\strut
\citet{Senft2008}
\strut\end{minipage} &
\begin{minipage}[t]{0.62\columnwidth}\raggedright\strut
IAPWS; \citet{Feistel2006}; \citet{Stewart2005}; BME
\strut\end{minipage}\tabularnewline
\begin{minipage}[t]{0.32\columnwidth}\raggedright\strut
\citet{Sotin2007}; \citet{Sotin2010}
\strut\end{minipage} &
\begin{minipage}[t]{0.62\columnwidth}\raggedright\strut
BME with thermal expansion (MGD)
\strut\end{minipage}\tabularnewline
\begin{minipage}[t]{0.32\columnwidth}\raggedright\strut
\citet{Valencia2006}
\strut\end{minipage} &
\begin{minipage}[t]{0.62\columnwidth}\raggedright\strut
BME with thermal expansion
\strut\end{minipage}\tabularnewline
\begin{minipage}[t]{0.32\columnwidth}\raggedright\strut
\citet{Valencia2010}
\strut\end{minipage} &
\begin{minipage}[t]{0.62\columnwidth}\raggedright\strut
\citet{French2009}; SESAME
\strut\end{minipage}\tabularnewline
\begin{minipage}[t]{0.32\columnwidth}\raggedright\strut
\citet{Vazan2013}
\strut\end{minipage} &
\begin{minipage}[t]{0.62\columnwidth}\raggedright\strut
Quotidian EOS; TFD
\strut\end{minipage}\tabularnewline
\begin{minipage}[t]{0.32\columnwidth}\raggedright\strut
\citet{Wilson2012}; \citet{Wilson2013}
\strut\end{minipage} &
\begin{minipage}[t]{0.62\columnwidth}\raggedright\strut
DFT
\strut\end{minipage}\tabularnewline
\begin{minipage}[t]{0.32\columnwidth}\raggedright\strut
\citet{Zeng2013}
\strut\end{minipage} &
\begin{minipage}[t]{0.62\columnwidth}\raggedright\strut
\citet{Frank2004}; \citet{French2009}; TFD
\strut\end{minipage}\tabularnewline
\bottomrule
\end{tabular}
\end{table}

\subsubsection{Our equation of state}\label{our-equation-of-state}

We extended the piecewise approach described above to include
temperature as a second dimension in parameter space. We drew from a
number of the sources of data listed in Table \ref{tbl:eos-sources}. Our
stitched equation of state is valid over a wide domain: its temperature
domain is from 275\(\,\)K to 24000\(\,\)K, and its pressure domain is
from \(10^5\,\)Pa (1\(\,\)bar) upwards. Our approach was similar to that
of \citet{Senft2008}, who generated a ``5-Phase'' equation of state
across different liquid, vapour, and ice phases. However, their work
focused on the lower temperatures needed to model impact craters. We
have explicitly included much higher temperatures so as to capture the
behaviour of large super-Earth planets: we expect the cores of these to
reach thousands of Kelvin. \citet{Valencia2010} also constructed a
similar equation of state, though using only data from SESAME and the
IAPWS formulation.

\begin{figure*}
\centering
\includegraphics{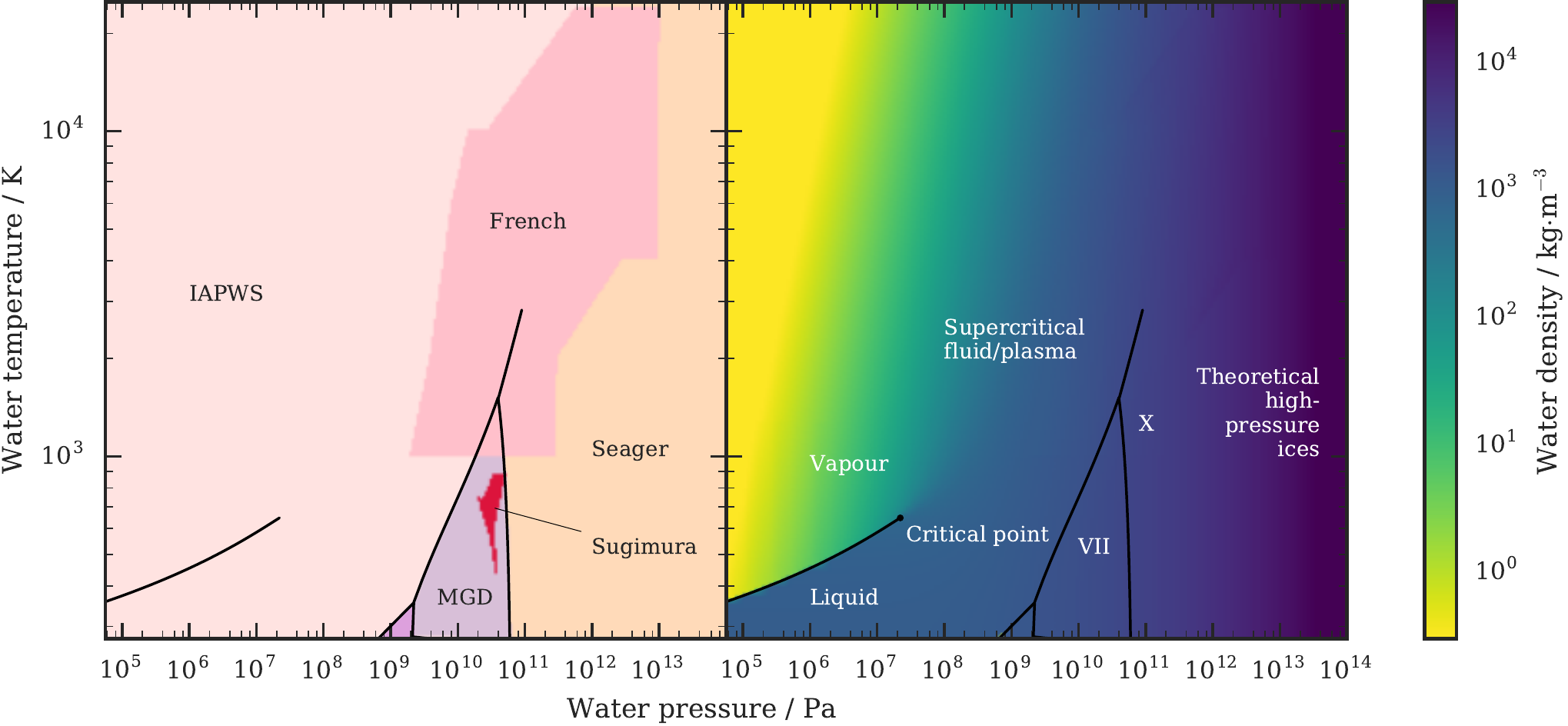}
\caption{Phases and data sources for our water equation of state. Our
equation of state covers a wide range of temperature--pressure space. On
the left, we show some of the our key data sources we used, and their
regions of validity: the IAPWS formulation by \citet{Wagner2002}; the
theoretical calculations of \citet{French2009}; the piecewise equation
of state described by \citet{Seager2007}; the Mie-Grüneisen-Debye (MGD)
thermal correction approach for ice VII in \citet{Sotin2007}; and the
measurements of \citet{Sugimura2010}, which cover a small region of ice
VII. We also show the relevant phase boundaries. On the right, we show
the density variation across the entire pressure--temperature range. The
density of water is more strongly affected by pressure across the range
we consider, but temperature also affects its density too, especially
across the liquid--vapour phase boundary and in the supercritical
region.\label{fig:eos-phase-space}}
\end{figure*}

\begin{figure}
\centering
\includegraphics{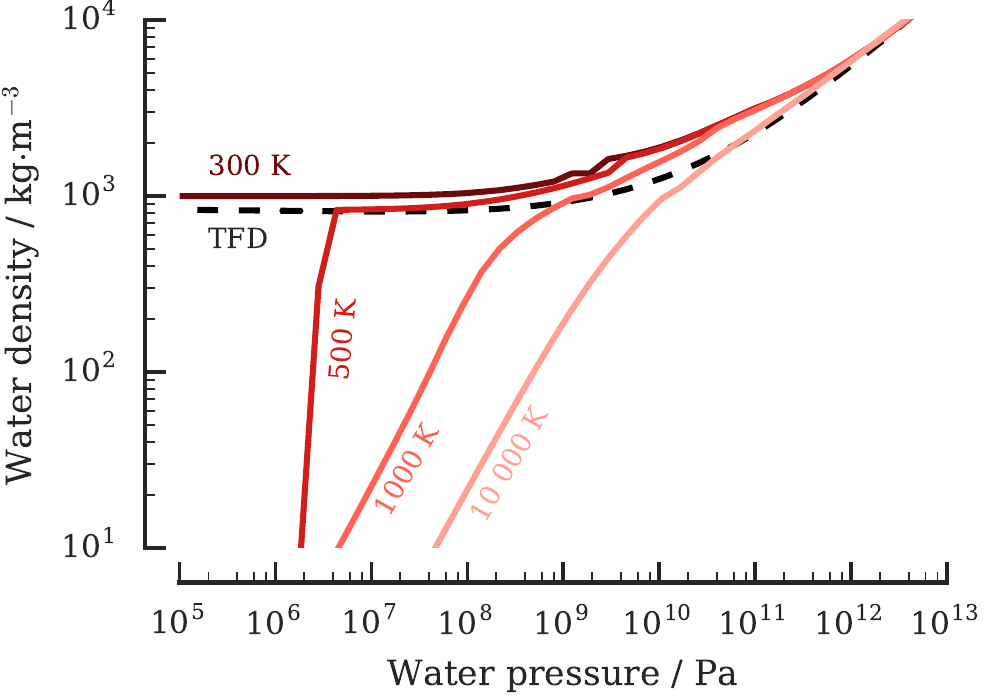}
\caption{Comparison of our equation of state with the high-pressure
limit. The TFD (Thomas--Fermi--Dirac) equation of state is increasingly
accurate in the high-pressure limit, where temperature effects on the
water density disappear. We also show temperature contours of our water
equation of state. The TFD, which has no temperature correction, is a
poor approximation of the behaviour of water at low pressures,
especially across the liquid--vapour phase boundary (vertical lines).
But all other choices of equation of state approach the TFD at high
pressures, and so it is appropriate in the TPa region and
beyond.\label{fig:eos-contours}}
\end{figure}

We endeavoured to choose equations of state that were most
representative of the thermal behaviour of water across this temperature
and pressure domain. We were guided by two principles in doing so.
First, as demonstrated in Fig. \ref{fig:eos-contours}, we expect thermal
expansion effects to approach zero as the pressure increases: this is a
consequence of the equations of state approaching the high-pressure TFD
limit. There are significant temperature effects at lower pressures, and
it is these effects we expected to be most important in our study.
Secondly, we aimed for a full treatment of density changes over phase
boundaries. Accordingly, we used the phase boundaries specified by
\citet{Dunaeva2010} to divide the temperature--pressure phase space into
regions corresponding to different phases of water. We then chose an
appropriate equation of state to represent each phase.

Our equation of state is for pure water only. Others have investigated
how impurities may affect the equation of state and the planet's
properties. For example, \citet{Levi2014} included a methane component
in their models, resulting in a new phase of water (filled ice) which
changes the planet's thermal profile. They note that, while neglecting
volatiles is an impediment to understanding the planet's atmosphere,
pure water models may be sufficient for planetary mass--radius
relations.

In selecting the equations of state we were often faced with choices
between different sources of data. The exact behaviour of water at very
high pressures is still uncertain and experimental and theoretical
results are sometimes in conflict \citep{Baraffe2014}. Ensuring absolute
accuracy of the chosen equations of state was therefore a secondary
priority. In general, we preferred more recent data to older data, we
prioritised measured and tabulated values over functional
approximations, and we chose representations that included temperature
dependence over those that did not. In the following paragraphs, we
describe our equation of state choices and summarise them in Table
\ref{tbl:our-eos}.

\begin{table}
\centering
\caption{We used a variety of equations of state in our final models.
``Tabular'' indicates that we interpolated between values specified in
the paper. ``Functional'' indicates that we used the functional form
given in the paper.\label{tbl:our-eos}}
\begin{tabular}{@{}lll@{}}
\toprule
\begin{minipage}[b]{0.29\columnwidth}\raggedright\strut
Equation of state
\strut\end{minipage} &
\begin{minipage}[b]{0.14\columnwidth}\raggedright\strut
Type
\strut\end{minipage} &
\begin{minipage}[b]{0.49\columnwidth}\raggedright\strut
Region of validity
\strut\end{minipage}\tabularnewline

\midrule
\begin{minipage}[t]{0.29\columnwidth}\raggedright\strut
IAPWS
\strut\end{minipage} &
\begin{minipage}[t]{0.14\columnwidth}\raggedright\strut
Tabular
\strut\end{minipage} &
\begin{minipage}[t]{0.49\columnwidth}\raggedright\strut
Vapour and liquid phases; 0.05 to 1000\(\,\)MPa and 252.462 to
1273\(\,\)K
\strut\end{minipage}\tabularnewline
\begin{minipage}[t]{0.29\columnwidth}\raggedright\strut
\citet{French2009}
\strut\end{minipage} &
\begin{minipage}[t]{0.14\columnwidth}\raggedright\strut
Tabular
\strut\end{minipage} &
\begin{minipage}[t]{0.49\columnwidth}\raggedright\strut
Superionic, plasma and high-pressure ice phases; 79 to
\(9.87×10^6\,\)MPa and 1000 to 24000\(\,\)K. We did not use table VIII
from this work, as this low-density data disagrees with the IAPWS
formulation.
\strut\end{minipage}\tabularnewline
\begin{minipage}[t]{0.29\columnwidth}\raggedright\strut
\citet{Feistel2006}
\strut\end{minipage} &
\begin{minipage}[t]{0.14\columnwidth}\raggedright\strut
Tabular
\strut\end{minipage} &
\begin{minipage}[t]{0.49\columnwidth}\raggedright\strut
Ice Ih; 0 to 200\(\,\)MPa and 0 to 273\(\,\)K
\strut\end{minipage}\tabularnewline
\begin{minipage}[t]{0.29\columnwidth}\raggedright\strut
\citet{Sugimura2010}
\strut\end{minipage} &
\begin{minipage}[t]{0.14\columnwidth}\raggedright\strut
Tabular
\strut\end{minipage} &
\begin{minipage}[t]{0.49\columnwidth}\raggedright\strut
Ice VII; 18880 to 50250\(\,\)MPa and 431 to 881\(\,\)K
\strut\end{minipage}\tabularnewline
\begin{minipage}[t]{0.29\columnwidth}\raggedright\strut
Vinet + MGD correction using parameters from \citet{Fei1993}
\strut\end{minipage} &
\begin{minipage}[t]{0.14\columnwidth}\raggedright\strut
Functional
\strut\end{minipage} &
\begin{minipage}[t]{0.49\columnwidth}\raggedright\strut
Ice VII
\strut\end{minipage}\tabularnewline
\begin{minipage}[t]{0.29\columnwidth}\raggedright\strut
TFD
\strut\end{minipage} &
\begin{minipage}[t]{0.14\columnwidth}\raggedright\strut
Functional
\strut\end{minipage} &
\begin{minipage}[t]{0.49\columnwidth}\raggedright\strut
Ice X
\strut\end{minipage}\tabularnewline
\begin{minipage}[t]{0.29\columnwidth}\raggedright\strut
\citet{Seager2007}
\strut\end{minipage} &
\begin{minipage}[t]{0.14\columnwidth}\raggedright\strut
Functional
\strut\end{minipage} &
\begin{minipage}[t]{0.49\columnwidth}\raggedright\strut
Ice VIII--X transition
\strut\end{minipage}\tabularnewline
\begin{minipage}[t]{0.29\columnwidth}\raggedright\strut
\citet{Choukroun2007}
\strut\end{minipage} &
\begin{minipage}[t]{0.14\columnwidth}\raggedright\strut
Functional
\strut\end{minipage} &
\begin{minipage}[t]{0.49\columnwidth}\raggedright\strut
Ices I, III, V, VI; phase boundaries as specified by \citet{Dunaeva2010}
\strut\end{minipage}\tabularnewline
\begin{minipage}[t]{0.29\columnwidth}\raggedright\strut
IAPWS extrapolations
\strut\end{minipage} &
\begin{minipage}[t]{0.14\columnwidth}\raggedright\strut
Functional
\strut\end{minipage} &
\begin{minipage}[t]{0.49\columnwidth}\raggedright\strut
Remaining regions
\strut\end{minipage}\tabularnewline
\bottomrule
\end{tabular}
\end{table}

\paragraph*{Liquid and vapour:}\label{liquid-and-vapour}
\addcontentsline{toc}{paragraph}{Liquid and vapour:}

The behaviour of water in the liquid and vapour phases is well
understood and there are plenty of data available. We were unable to
gain access to the SESAME tables of \citet{Lyon1992} because there are
restrictions on the distribution of this data to non-US nationals.
Instead, to represent water liquid and vapour, we selected the IAPWS
(International Association for the Properties of Water and Steam)
formulation \citep{Wagner2002}, which provides both tabular and
functional data for water in these phases. These are well-tested and
validated by years of experiments. \citet{Wagner2002} also claim that
the functional forms can be extrapolated outside the range of the
tables. We implemented the functional relationships between temperature,
density and pressure. Where appropriate, we numerically inverted these
to give a relation of the form \(ρ = ρ(P, T)\). We then tested these
against the tables to verify that we had replicated them correctly.

\paragraph*{Ice VII:}\label{ice-vii}
\addcontentsline{toc}{paragraph}{Ice VII:}

We explicitly chose a temperature-dependent formulation because we
expected ice VII to form a significant fraction of the planet in the
cases where the water layer is large. This temperature-dependent
formulation is in contrast to other studies which have assumed that the
ice VII layer is isothermal: for example, \citet{Rogers2010a} assumed no
expansion in all solid layers, choosing to include temperature effects
only in the gaseous and liquid phases.

The best temperature-dependent formulation we found for ice VII was the
Mie-Grüneisen-Debye (MGD) thermal correction approach described by
\citet{Sotin2007}. We used a Vinet equation of state with this thermal
correction, taking the coefficients of \citet{Fei1993}, within the ice
VII region delimited by the phase boundaries of \citet{Dunaeva2010}.
However, we preferred the more recent tabulated measurements of
\citet{Sugimura2010} wherever these were applicable; these are shown
within the ice VII region in Fig. \ref{fig:eos-phase-space}.

\paragraph*{Supercritical fluid and
plasma:}\label{supercritical-fluid-and-plasma}
\addcontentsline{toc}{paragraph}{Supercritical fluid and plasma:}

\citet{French2009} presented quantum molecular dynamics simulations of
high-temperature and high-pressure plasma, ice, and superionic fluid
phases of water. We used their tables in the region beyond 1000\(\,\)K
and \(1.86×10^9\,\)Pa. \citet{Lopez2012} noted that this region has
recently been probed by laboratory experiments thanks to
\citet{Knudson2012}, who strongly advocate ``that {[}the French equation
of state{]} be the standard in modeling water in Neptune, Uranus, and
`hot Neptune' exoplanets.''. These temperatures and pressures are also
relevant to the interiors of super-Earths. We did not use the
low-density tables that they presented separately because these differ
significantly from the IAPWS results in the same temperature and
pressure range.

\paragraph*{Low-temperature ices:}\label{low-temperature-ices}
\addcontentsline{toc}{paragraph}{Low-temperature ices:}

For completeness, our equation of state includes low-pressure ice Ih
from \citet{Feistel2006} as well as higher-pressure ices such as ice
III, V and VI. We took the phase boundaries from \citet{Dunaeva2010} and
used the temperature-dependent formulations for these ices by
\citet{Choukroun2007}.

\paragraph*{Ice X and beyond:}\label{ice-x-and-beyond}
\addcontentsline{toc}{paragraph}{Ice X and beyond:}

We adopted the piecewise equation of state of \citet{Seager2007} to
describe the transition from ice VII to ice X and beyond. This does not
include any temperature dependence: any interesting phase behaviour of
ice at these high pressures is increasingly theoretical and unconfirmed
by experiment. Temperature effects approach zero at these high pressures
anyway (Fig. \ref{fig:eos-contours}), so we used the
Thomas--Fermi--Dirac equation of state for all regions beyond 7686GPa
which were not covered by one of the other regions listed above.

\paragraph*{Other regions:}\label{other-regions}
\addcontentsline{toc}{paragraph}{Other regions:}

Finally, we filled in all other regions according to the IAPWS
formulation or extrapolations thereof. In practice, the only regions not
covered above were low-pressure and high-temperature vapour regions,
which we do not expect to be relevant for our super-Earth interior
models.

\begin{table}
\centering
\caption{Sources for the abbreviated equation of state designations used
in this paper.\label{tbl:eos-abbrevs}}
\begin{tabular}{@{}ll@{}}
\toprule
\begin{minipage}[b]{0.24\columnwidth}\raggedright\strut
Equation of state
\strut\end{minipage} &
\begin{minipage}[b]{0.71\columnwidth}\raggedright\strut
Source
\strut\end{minipage}\tabularnewline

\midrule
\begin{minipage}[t]{0.24\columnwidth}\raggedright\strut
ANEOS
\strut\end{minipage} &
\begin{minipage}[t]{0.71\columnwidth}\raggedright\strut
\citet{Thompson1972}
\strut\end{minipage}\tabularnewline
\begin{minipage}[t]{0.24\columnwidth}\raggedright\strut
BME
\strut\end{minipage} &
\begin{minipage}[t]{0.71\columnwidth}\raggedright\strut
Birch--Murnaghan equation of state; see \citet{Poirier2000}
\strut\end{minipage}\tabularnewline
\begin{minipage}[t]{0.24\columnwidth}\raggedright\strut
DFT
\strut\end{minipage} &
\begin{minipage}[t]{0.71\columnwidth}\raggedright\strut
Density functional theory; refers to theoretical calculations which
multiple authors have performed
\strut\end{minipage}\tabularnewline
\begin{minipage}[t]{0.24\columnwidth}\raggedright\strut
H\(_2\)O-REOS
\strut\end{minipage} &
\begin{minipage}[t]{0.71\columnwidth}\raggedright\strut
\citet{Nettelmann2011}; includes IAPWS, SESAME, \citet{French2009},
\citet{Feistel2006}
\strut\end{minipage}\tabularnewline
\begin{minipage}[t]{0.24\columnwidth}\raggedright\strut
IAPWS
\strut\end{minipage} &
\begin{minipage}[t]{0.71\columnwidth}\raggedright\strut
\citet{Wagner2002}
\strut\end{minipage}\tabularnewline
\begin{minipage}[t]{0.24\columnwidth}\raggedright\strut
LM-REOS
\strut\end{minipage} &
\begin{minipage}[t]{0.71\columnwidth}\raggedright\strut
\citet{Nettelmann2008} (precursor to H\(_2\)O-REOS)
\strut\end{minipage}\tabularnewline
\begin{minipage}[t]{0.24\columnwidth}\raggedright\strut
MGD
\strut\end{minipage} &
\begin{minipage}[t]{0.71\columnwidth}\raggedright\strut
Mie-Grüneisen-Debye thermal pressure expansion; described in
\citet{Sotin2007}
\strut\end{minipage}\tabularnewline
\begin{minipage}[t]{0.24\columnwidth}\raggedright\strut
SESAME
\strut\end{minipage} &
\begin{minipage}[t]{0.71\columnwidth}\raggedright\strut
\citet{Lyon1992}
\strut\end{minipage}\tabularnewline
\begin{minipage}[t]{0.24\columnwidth}\raggedright\strut
TFD
\strut\end{minipage} &
\begin{minipage}[t]{0.71\columnwidth}\raggedright\strut
Thomas--Fermi--Dirac; described in \citet{Salpeter1967}
\strut\end{minipage}\tabularnewline
\begin{minipage}[t]{0.24\columnwidth}\raggedright\strut
Vinet
\strut\end{minipage} &
\begin{minipage}[t]{0.71\columnwidth}\raggedright\strut
\citet{Vinet1987}
\strut\end{minipage}\tabularnewline
\bottomrule
\end{tabular}
\end{table}

\subsubsection{Dealing with fragmented
data}\label{dealing-with-fragmented-data}

We made no attempt to smooth or otherwise interpolate between the
different sources of data described above. This approach means that
sharp density changes across phase boundaries are well-represented in
the final equation of state. This is desirable so that we may examine
the differentiation that results solely from phase transitions. It also
results in some artificial density discontinuities at the boundaries
between different data sets. We believe that this has not affected the
results: these discontinuities are minor compared with the density
variations within each phase of water and, in most cases, we also
bounded the domain of each data set to that of a particular phase.

Because we used disparate sources of data, we evaluated the density at a
given temperature and pressure in different ways depending on the data
source. Although we did not smooth or interpolate \emph{between} data
sets, we needed to interpolate some data sources \emph{within} the data
set. Where data were published in tabulated form on a structured grid,
we used simple two-dimensional linear interpolation\footnote{For
  multidimensional linear interpolation we used
  \href{https://github.com/kbarbary/Dierckx.jl}{Dierckx.jl}.} to
evaluate the equation of state at points not lying on the grid. Where
data were published as unstructured points, we used barycentric
interpolation on the mesh of Delaunay triangles\footnote{To construct a
  Delaunay tessellation we used
  \href{https://github.com/JuliaGeometry/VoronoiDelaunay.jl}{VoronoiDelaunay.jl}.}
defined by these points. We also used this Delaunay mesh to determine if
a given \((P,T)\) pair lay within the domain of a particular equation of
state, allowing us to fall back to another equation of state if
necessary. We evaluated functional forms of the equation of state as is,
defining their domain by means of a bounding box or a polygon in
\((P,T)\) space taken from the phase boundaries of \citet{Dunaeva2010}.

Some of the equations of state used in this final synthesized version
were much simpler than others. This meant that the evaluation time
varied from point to point, from very quick table lookups and
interpolation to the slower IAPWS formulae. In addition, any equation of
state that was specified in the inverse form \(P = P(ρ, T)\) needed to
be numerically inverted to give the canonical form \(ρ = ρ(P, T)\) used
in our models. To avoid duplicating this calculation unnecessarily, we
re-sampled the final equation of state on to a 256 by 256
pressure--temperature grid. Pre-computing and tabulating the data in
this way saved significant time. In our trials, the resolution of the
grid barely altered the properties of the planetary models. This
suggests that the density behaviour within a single phase region was
more important than any effects across phase boundaries that might be
lost by sampling from this discrete grid.

The equation of state we used necessarily has some uncertainty in it,
especially in regions near the critical point of water
\citep{Wagner2002} and at high temperatures and pressures where there
are sometimes conflicting experimental and theoretical data
\citep{Baraffe2008}. The error in the equation of state varies depending
on the original data source. For the region encompassed by the IAPWS
data \citep{Wagner2002}, the density uncertainty is approximately 0.01
per cent (liquid and solid), 0.03 to 0.1 per cent (vapour), and up to
0.5 per cent in the region around and beyond the critical point. They
give a more detailed breakdown of these errors in their section 6.3.2,
in particular fig.~6.1. We estimate that the error beyond these regions
is closer to 1 per cent if we extrapolate beyond the table and assume
that the uncertainty continues to increase at higher temperatures and
pressures. For the supercritical fluid, plasma and superionic phases in
the data of \citet{French2009}, they state that ``the QMD EOS is
accurate up to 1 per cent for the conditions relevant for the giant
planet's interiors of our solar system.'' For the ice VII phase, the
measurements of \citet{Sugimura2010} have errors of between 0.003 per
cent and 0.5 per cent. Finally, it is not possible to give a meaningful
uncertainty estimate at higher pressures where no measurements exist,
but we do not treat the temperature dependence there anyway.

\subsubsection{Thermal expansion and heat
capacity}\label{thermal-expansion-and-heat-capacity}

Equation \ref{eq:temp-gradient} requires both a heat capacity \(c_p\)
and a thermal expansion coefficient \(α\) (defined in equation
\ref{eq:thermal-expansion}). Following our goal of handling temperature
effects appropriately, we explicitly sought out temperature-dependent
forms for these.

We used the IAPWS tables for heat capacity in the liquid--vapour range,
then took the nearest available data point from these tables for all
other pressure--temperature points. This is because we could not find
readily available heat capacity data across the full range of phases in
our equation of state. This approach therefore does not reflect any
changes in heat capacity between the high-pressure ice phases. The most
significant effect is the change in heat capacity across the
liquid--vapour phase boundary, which we do capture in our models.

We drew the thermal expansion coefficient \(α\) directly from the
equation of state by evaluating equation \ref{eq:thermal-expansion}. We
used automatic differentiation\footnote{We used forward-mode automatic
  differentiation provided by
  \href{https://github.com/JuliaDiff/ForwardDiff.jl}{ForwardDiff.jl}.}
where possible to evaluate the derivative. In some cases this was not
possible\footnote{The Delaunay triangulation method in the library we
  used incorporates a method called floating-point filtering, which
  relies on the specific properties of floating point numbers. It could
  not be used with the automatic differentiation approach we used, which
  evaluates functions as usual but replaces the inputs with a special
  numeric type.} so we used finite differencing\footnote{We used the
  package
  \href{https://github.com/johnmyleswhite/Calculus.jl}{Calculus.jl}.}.
As well as pre-computing the equation of state itself, we pre-computed
and tabulated the thermal expansion coefficient on the same
pressure--temperature grid. Some previous works have assumed a fixed
thermal expansion coefficient: for example, \citet{Ehrenreich2006} took
a fixed value for \(α\) in their models. We believe that our approach is
more appropriate for understanding how the temperature gradient and
physical properties of a watery planet are affected by the thermal
properties of water.

\subsection{Model verification}\label{model-verification}

We verified our models by making mass--radius diagrams as described in
the previous section and comparing them with previous work.

\subsubsection{The isothermal case}\label{the-isothermal-case}

\begin{figure}
\centering
\includegraphics{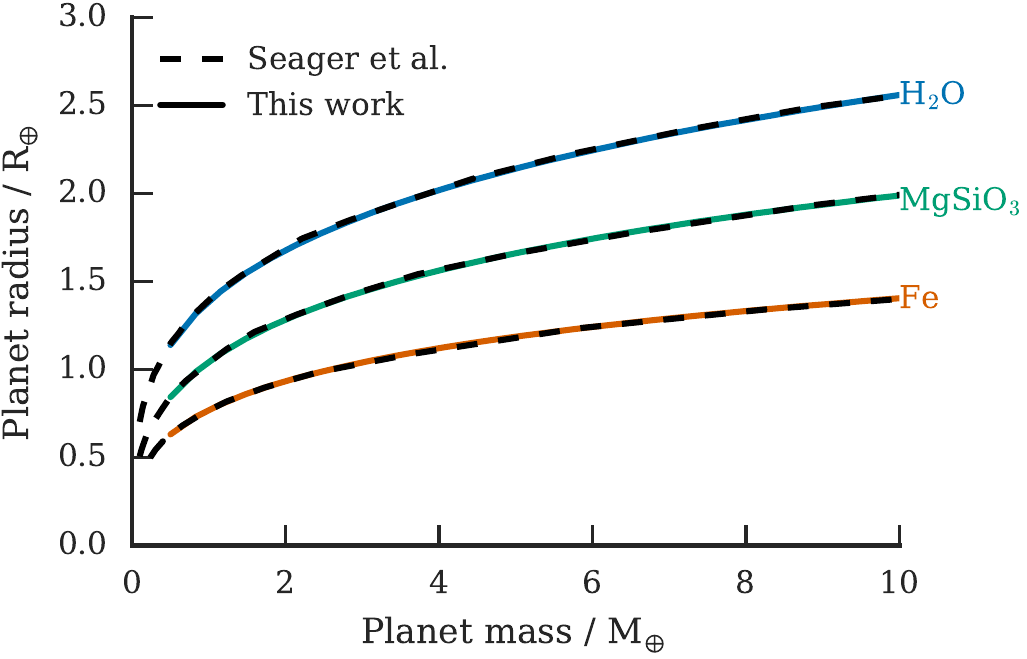}
\caption{Validation of isothermal models. Our structural models exactly
reproduce previous results in the isothermal case. Here we show
mass--radius relations for homogeneous isothermal spheres. If we adopt
identical equations of state to those used by \citet{Seager2007}, we
obtain the same result. This serves as a verification that we are
correctly solving the structural equations. These models used zero
surface pressure and have no temperature dependence: the equations of
state are isothermal and are taken at
300\(\,\)K.\label{fig:seager-mr-comparison}}
\end{figure}

We checked that our models work in the isothermal case by replicating
the mass--radius relations of \citet{Seager2007}. We exactly reproduced
the mass--radius relations when we constructed homogeneous isothermal
300\(\,\)K planets using the equations of state specified in their
paper, as shown in Fig. \ref{fig:seager-mr-comparison}. We set the
surface pressure of our models to zero, following the boundary condition
they used. The surface pressure hardly affects the results because the
equations of state are for the solid phase only. This identical
mass--radius relation verified that our integrator works correctly, and
we therefore began to investigate where the differences lie upon
including temperature effects.

\subsubsection{The adiabatic case}\label{the-adiabatic-case}

\begin{figure}
\centering
\includegraphics{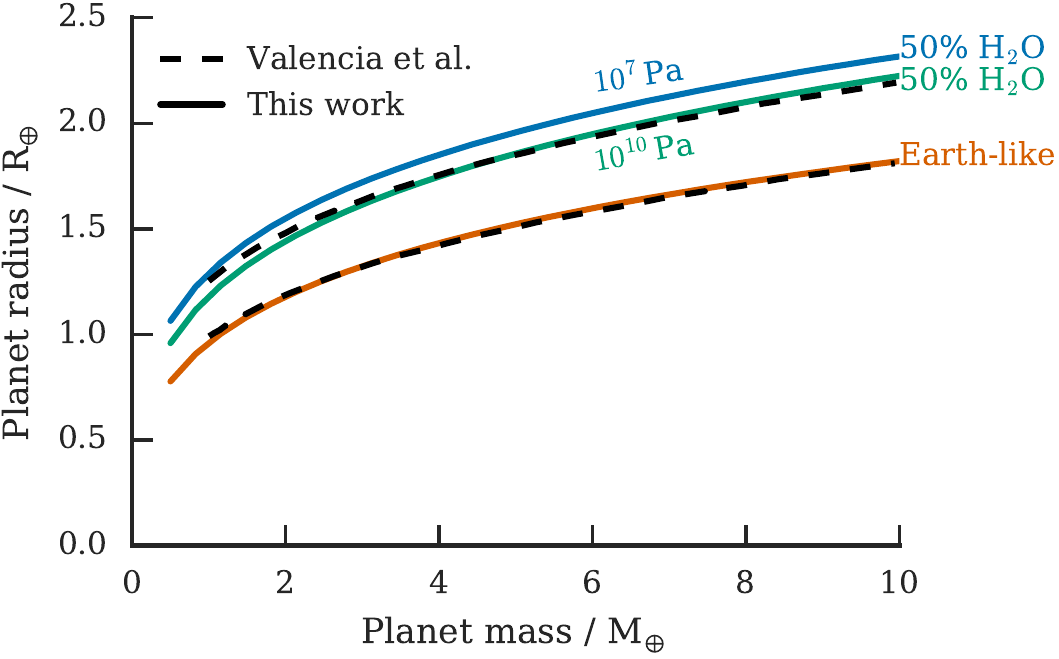}
\caption{Validation of adiabatic models. Our mass--radius relations
reproduce those for dry planets well, and predict inflated radii for
planets with water layers. Here we show mass--radius relations for two
classes of models: dry planets (33 per cent Fe and 67 per cent
MgSiO\(_3\) by mass), and wet planets (17 per cent Fe, 33 per cent
MgSiO\(_3\), and 50 per cent water). We compared the mass--radius
relations with the work of \citet{Valencia2007a} who constructed models
with ice VII layers. At a surface pressure of \(10^{10}\,\)Pa the water
layer in the wet planets is mostly ice VII and so our results are
similar in this case. Small differences are likely due to our different
equation of state choice for ice VII. However, at lower surface
pressures, water can have an extended lower density shell that results
in a larger planet than otherwise expected. The surface temperature in
these models is 550\(\,\)K, matching the characteristic temperature used
by \citet{Valencia2007a} in their
models.\label{fig:valencia-mr-comparison}}
\end{figure}

We verified our adiabatic multi-layer models by comparing them with
those of \citet{Valencia2007a}, who constructed similar models using the
ice VII equation of state for water (Fig.
\ref{fig:valencia-mr-comparison}). When we set high surface pressures
(\(10^{10}\,\)Pa) we forced the surface layer of water to begin as ice
VII or close to it and therefore produced a very similar mass--radius
relation. However, we predict inflated radii at lower surface pressures
and therefore conclude that surface temperature and surface pressure are
both important factors for determining the radius of a planet with a
water layer. We further explore this relationship in our results
section.

There are minor differences between our mass--radius relations and the
mass--radius relations presented by \citet{Valencia2007a}. We slightly
underpredict the radii of lower-mass planets in models with surface
pressures of \(10^{10}\,\)Pa. These differences are likely due to our
choice of equation of state: we use only simple isothermal prescriptions
for iron and magnesium silicate and include more phases of water than
just ice VII. We also did not include any treatment of conductive
boundary layers in our models. In general, however, our results agree
well with theirs.

\begin{figure}
\centering
\includegraphics{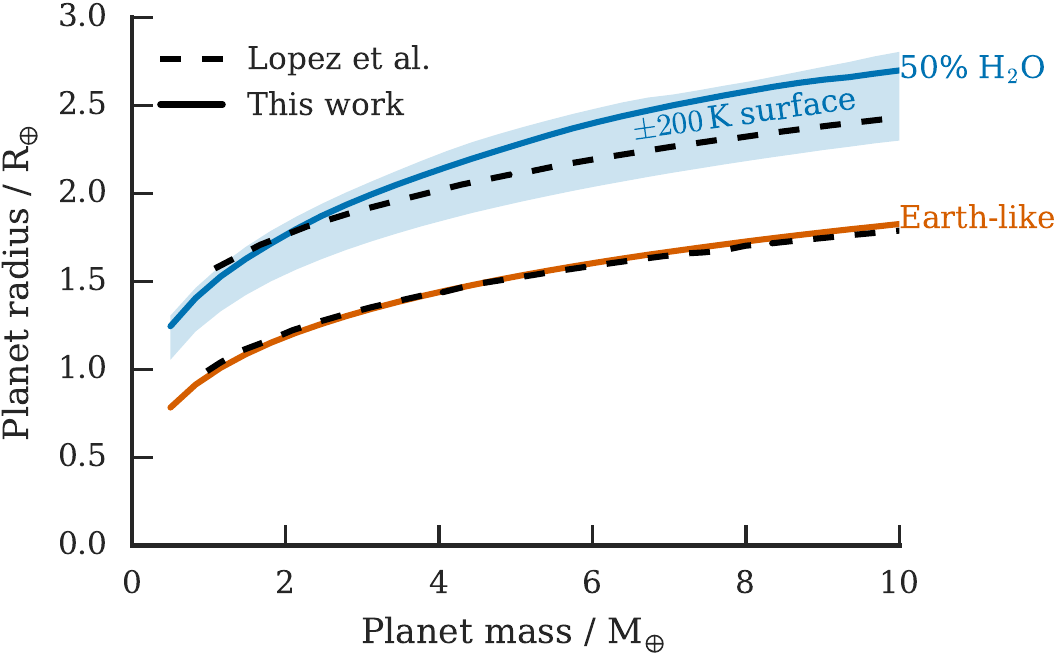}
\caption{Comparison with evolutionary models. We plot dry (Earth-like)
and wet (50 per cent water on an Earth-ratio core/mantle) mass--radius
relations. Shown for comparison are models by \citet{Lopez2012}, who
build on work by \citet{Fortney2007} and \citet{Nettelmann2011} by using
a thermal evolution approach to track the entropy within each planet as
it cools. Surface temperature significantly alters the mass--radius
relation in our models. The surface temperature in these models is
700\(\,\)K but the shaded band shows models with surface temperatures
from 500 to 900\(\,\)K, a significant spread, which is caused by
temperature-dependent density changes of water at lower pressures. We
chose a surface pressure of \(10^7\,\)Pa to approximately match the
radii of \citet{Lopez2012}. Their method does not begin from an explicit
surface pressure, as ours does.\label{fig:lopez-mr-comparison}}
\end{figure}

We also compared our results with the evolutionary models of
\citet{Lopez2012}. Although we were able to reproduce their mass--radius
relation for Earth-like planets, we were less successful when adding
extended water layers (Fig. \ref{fig:lopez-mr-comparison}). We can match
the radius of an arbitrary planet by choosing an appropriate surface
pressure but we underpredict the radii of small planets and overpredict
the radii of large planets compared with their results. This may be a
result of different equation of state choices or different temperature
gradients during the course of their evolutionary calculations.

Fig. \ref{fig:lopez-mr-comparison} also provides a first indication of
how changes in surface temperature can affect the mass--radius relation.
We highlight the magnitude of these differences and note that they are
still significant at pressures of \(10^7\,\)Pa (100\(\,\)bar) and up,
well into the pressure region where many atmospheric models terminate.
We explore the effects on our models of changing surface temperature,
surface pressure and composition in the next section.

\section{Results}\label{results}

We have explored the effects of temperature dependence on the radii of
water-rich super-Earths. This section shows that significant radius
variations can occur across temperature ranges relevant to super-Earths.
We explored the dependence of super-Earth radii on three key model
parameters.

\begin{enumerate}
\def\labelenumi{\arabic{enumi}.}
\tightlist
\item
  Planet surface temperature, with the water layer temperature profile
  taken as

  \begin{enumerate}
  \def\labelenumii{\alph{enumii})}
  \tightlist
  \item
    isothermal, or
  \item
    adiabatic.
  \end{enumerate}
\item
  Planet surface pressure.
\item
  Planet composition, i.e.~water mass fraction.
\end{enumerate}

\subsection{Effect of surface temperature on isothermal and adiabatic
interiors}\label{effect-of-surface-temperature-on-isothermal-and-adiabatic-interiors}

We found that thermal expansion can lead to significant changes in the
radii of water-rich super-Earths. We constructed super-Earths in two
different ways. First we modelled them as isothermal spheres containing
an Earth-like core (33 per cent Fe and 67 per cent MgSiO\(_3\))
underneath a water layer of 30 per cent of the planet's mass. Then we
instead allowed the temperature to increase adiabatically into the water
layer. Fig. \ref{fig:isothermal-vs-adiabatic} shows that the assumption
that thermal expansion effects are negligible, which was made in some
previous studies, is not the case. This is true in two senses. First, a
significant temperature dependence exists when we adopt an adiabatic
interior temperature profile compared with an isothermal one. The
surface temperature also affects the radius of a planet within both
types of models.

\begin{figure}
\centering
\includegraphics{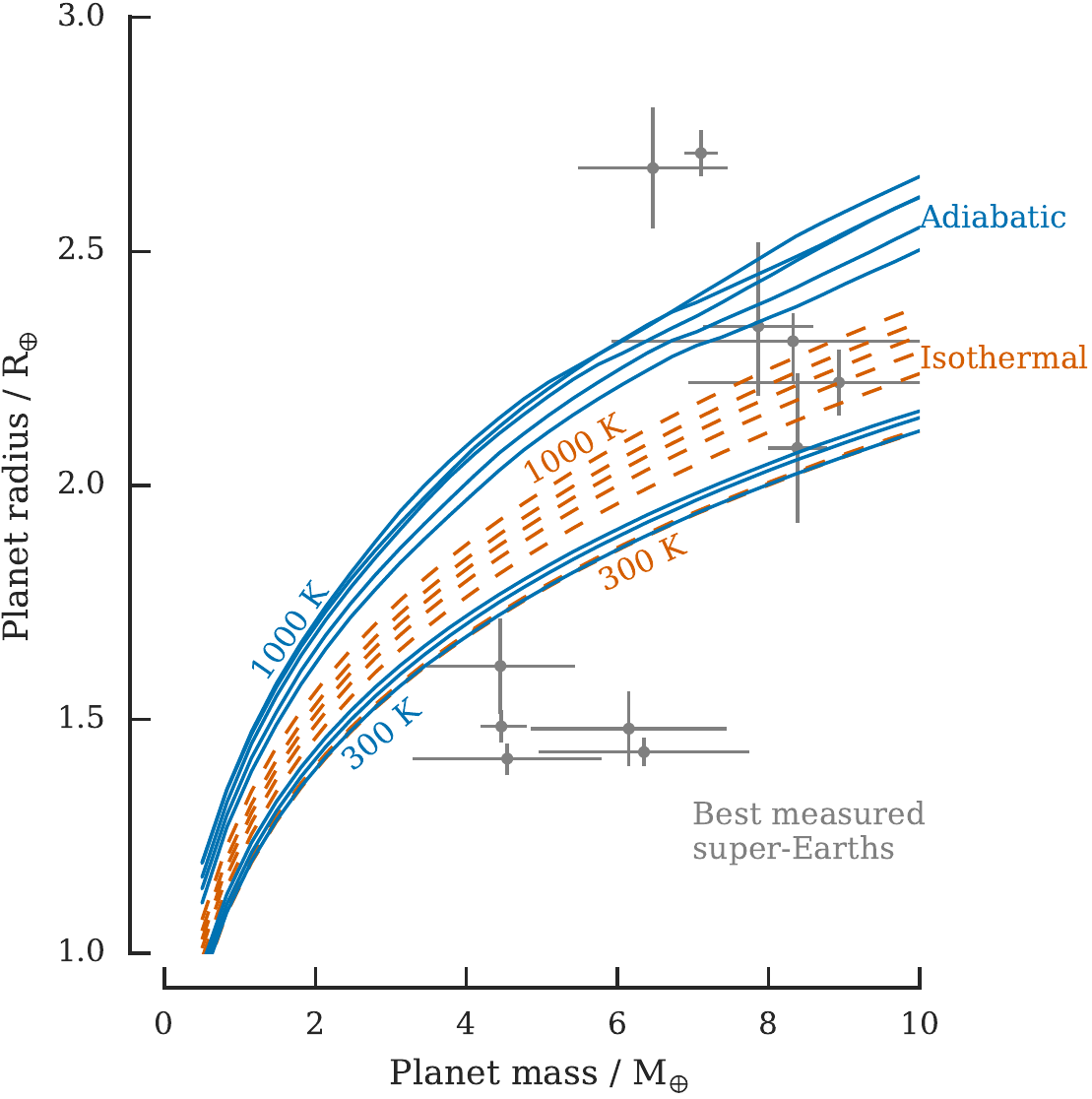}
\caption[Dependence of watery super-Earth radii on surface temperature
and internal temperature profile. An increased surface temperature
results in an increased planetary radius. This effect is especially
pronounced in the full adiabatic temperature treatment. Here we show
super-Earths with an Earth-like core under a 30 per cent water layer by
mass. We treated the temperature in two different ways: an isothermal
treatment with a fixed constant temperature and an adiabatic treatment
where we fixed the surface temperature but allowed the temperature to
increase inwards according to the adiabatic relation (equation
\ref{eq:temp-gradient}). The adiabatic models are warmer and therefore
significantly larger overall, but even the isothermal planets display
some radius change due to temperature. The effects of this temperature
dependence are comparable to current uncertainties on measured masses
and radii for some of the best-characterised exoplanets. The surface
pressure in these models is \(10^7\,\)Pa (100\(\,\)bar), and the
temperature increases in steps of 100\(\,\)K. The large gap between 500
and 600\(\,\)K in the adiabatic case is due to a density discontinuity
between the liquid and vapour
phases.\label{fig:isothermal-vs-adiabatic}]{Dependence of watery
super-Earth radii on surface temperature and internal temperature
profile. An increased surface temperature results in an increased
planetary radius. This effect is especially pronounced in the full
adiabatic temperature treatment. Here we show super-Earths with an
Earth-like core under a 30 per cent water layer by mass. We treated the
temperature in two different ways: an isothermal treatment with a fixed
constant temperature and an adiabatic treatment where we fixed the
surface temperature but allowed the temperature to increase inwards
according to the adiabatic relation (equation \ref{eq:temp-gradient}).
The adiabatic models are warmer and therefore significantly larger
overall, but even the isothermal planets display some radius change due
to temperature. The effects of this temperature dependence are
comparable to current uncertainties on measured masses and radii for
some of the best-characterised exoplanets\footnotemark{}. The surface
pressure in these models is \(10^7\,\)Pa (100\(\,\)bar), and the
temperature increases in steps of 100\(\,\)K. The large gap between 500
and 600\(\,\)K in the adiabatic case is due to a density discontinuity
between the liquid and vapour
phases.\label{fig:isothermal-vs-adiabatic}}
\end{figure}
\footnotetext{These data are from
  \href{http://www.exoplanets.org}{exoplanets.org}. We selected planets
  with known radii and masses of 1 to 10\(\,\)M\(_⊕\). We then plotted
  the twelve planets with the lowest summed relative uncertainty in mass
  and radius \(\left( ΔR/R + ΔM/M \right)\).}

The adiabatic models have a larger radius for a given mass when compared
with the isothermal case. This is to be expected: the average
temperature is higher along an adiabat than an isotherm fixed at the
surface temperature, and the density of water generally decreases with
temperature. The increase in radius is significant at higher surface
temperatures, as shown in Fig. \ref{fig:isothermal-vs-adiabatic}. For
example, a 4\(\,\)M\(_⊕\) 30 per cent water planet with a 600\(\,\)K
surface has a radius of 1.8\(\,\)R\(_⊕\) if its water layer is
isothermal, but 2\(\,\)R\(_⊕\) if it is adiabatic. Across the
super-Earth mass range we considered, the adiabatic radii increased by
up to 0.3\(\,\)R\(_⊕\) when compared with the isothermal case. The
difference becomes particularly pronounced at higher surface
temperatures, at which point the water layer may consist of
supercritical fluid rather than liquid, solid, or vapour (Fig.
\ref{fig:water-phases}).

A significant dependence on surface temperature also exists when using
the adiabatic models. That is, changing the surface temperature affects
the radius of a model water super-Earth even when its temperature
profile is already being treated as adiabatic. In the case of a
10\(\,\)M\(_⊕\) planet, increasing the surface temperature from 300 to
1000\(\,\)K gave a radius increase of 0.6\(\,\)R\(_⊕\). For an
Earth-mass planet the increase was approximately 0.3\(\,\)R\(_⊕\) for
the same temperature range.

We have highlighted above the change in the adiabatic models, which we
claim are a more realistic representation of the actual temperature
structure within the planet. But even the isothermal models show a
significant increase in radius with the planet's temperature. For a
10\(\,\)M\(_⊕\) planet, the change in radius is 0.3\(\,\)R\(_⊕\) from
300 to 1000\(\,\)K. This is due to the thermal expansion of the planet
as a whole, rather than of one small part of the water layer near the
surface. We do not necessarily expect an adiabatic temperature gradient
throughout the whole planet because the entire interior may not all be
convective. For example, \citet{Valencia2007a} included conductive
boundary layers in their models. In that case, the true
temperature-dependent behaviour of the mass--radius diagram might lie
between the adiabatic and isothermal cases. Despite this, Fig.
\ref{fig:isothermal-vs-adiabatic} shows that the surface temperature can
still play an important role in determining the radius of a planet if it
has a substantial water layer. This is true even in the extreme
isothermal case where there is no temperature gradient at all within the
planet.

These models have a surface pressure of \(10^7\,\)Pa (100\(\,\)bar) so
this effect is not due to the strong liquid--vapour transition at
1\(\,\)bar. In fact, we still see these effects past the critical
pressure of water (\(2.206×10^7\,\)Pa). The critical point, which is
visible in Figs \ref{fig:water-phases} and \ref{fig:eos-phase-space}, is
the point in temperature--pressure space beyond which there is no
distinct phase transition from liquid to vapour. This indicates that a
liquid--vapour transition is not required to produce a significantly
inflated radius when the water layer is heated. We discuss the effect of
pressure on these models further in the next section.

\subsection{Effect of surface
pressure}\label{effect-of-surface-pressure}

\begin{figure*}
\centering
\includegraphics{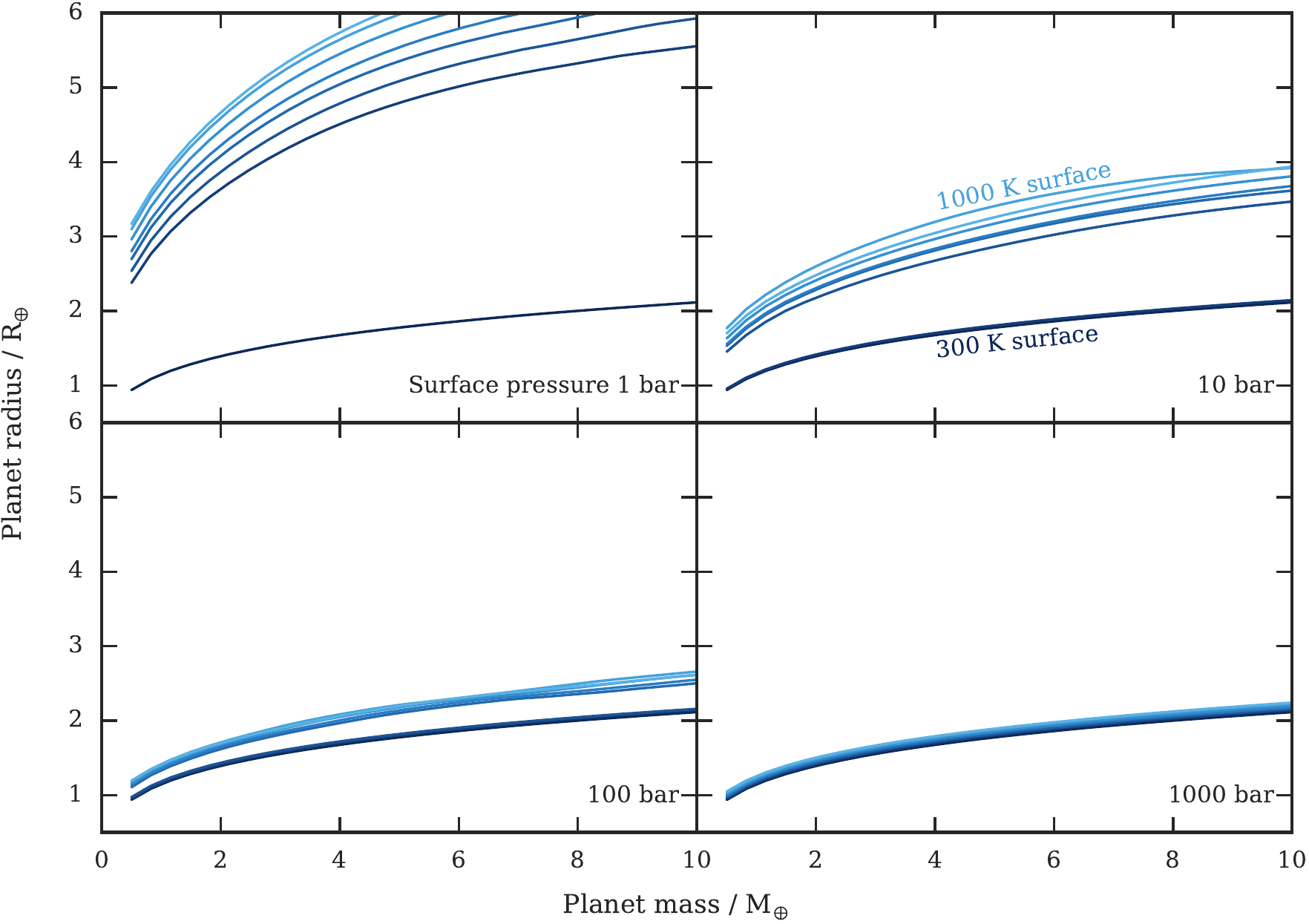}
\caption{Dependence of radii on surface pressure. The effect of
temperature on the radius of watery planets decreases with increasing
surface pressure, but remains significant (greater than about
\(0.1\,\)R\(_⊕\)) for pressures below 1000\(\,\)bar. Here we show
mass--radius relations for spheres with an Earth-like core under a 30
per cent water layer, changing only the surface pressure each time. The
temperature dependence remains even beyond the critical pressure of
water (\(2.206×10^7\,\)Pa), at which point the surface water exists as a
supercritical fluid. Only at very high pressures (\(10^9\) or
\(10^{10}\,\)Pa; \(10\,000\) or \(100\,000\) bar) does this temperature
dependence vanish.\label{fig:surface-pressure-variation}}
\end{figure*}

The surface pressure can strongly affect the temperature-dependent
thickness of the water layer (Fig.
\ref{fig:surface-pressure-variation}). For example, at high temperatures
(1000\(\,\)K), increasing the surface pressure of a 10 per cent water
and 4\(\,\)M\(_⊕\) planet from 10\(\,\)bar to 1000\(\,\)bar compresses
the water layer significantly, decreasing the planet's radius by a
factor of two. And at low pressures we see a bifurcation in the surface
pressure contours where a surface temperature increase of 100\(\,\)K or
less can inflate the radius of a watery super-Earth by more than 50 per
cent. This is the result of a transition across the liquid--vapour phase
boundary, which exists at pressures up to the critical pressure of water
(\(2.206×10^7\,\)Pa). Our interior structure code is most likely not the
best choice for modelling such a quasi-atmospheric layer: we did not
handle radiative energy transfer in our models. We have therefore not
undertaken a detailed study of the behaviour of these vapour layers.
They likely require a more sophisticated treatment of the temperature
profile than our adiabatic assumption.

Despite observing highly inflated radii when the temperature is
increased across the liquid--vapour phase boundary, we still see
temperature-dependent variation in the planet's radius past the critical
pressure of water. This is because the density of water is still
strongly temperature-dependent in the super-critical regime. In fact, we
might reasonably expect the same inflated radii in any situation where
the pressure of the water layer places it in a region of the water phase
diagram that has significant temperature dependence. If the water layer
is heated to thousands of Kelvin, this temperature dependence may only
begin to disappear around \(10^{10}\,\)Pa (\(100\,000\,\)bar, Fig.
\ref{fig:eos-contours}). At a pressure of \(10^8\,\)Pa (1000\(\,\)bar),
a watery super-Earth with a surface temperature of 1000\(\,\)K still has
a radius that is up to 0.1\(\,\)R\(_⊕\) larger than one with a surface
temperature of 300\(\,\)K. This is comparable to or greater than the
best current uncertainties on measured super-Earth radii (Fig.
\ref{fig:isothermal-vs-adiabatic}), and indicates that the surface
temperature is a key parameter to consider when one attempts to model
planets with significant water mass.

We included no atmospheric layers in these models. Other studies have
provided more complete treatments of atmospheric layers. For example,
\citet{Rogers2010a} included a gas layer on top of an isothermal
interior structure model in order to interpret the structure of the
planet GJ 1214b. And \citet{Valencia2013} used internal structure models
coupled with an atmospheric layer, exploring the dependence of radii on
various model parameters including equilibrium temperature and water
content. Given that we set the surface pressure to between \(10^5\) and
\(10^{10}\,\)Pa (1 and \(100\,000\,\)bar), our models must therefore
represent the layers interior to an atmosphere of some sort.

\subsection{Effect of water content}\label{effect-of-water-content}

\begin{figure*}
\centering
\includegraphics{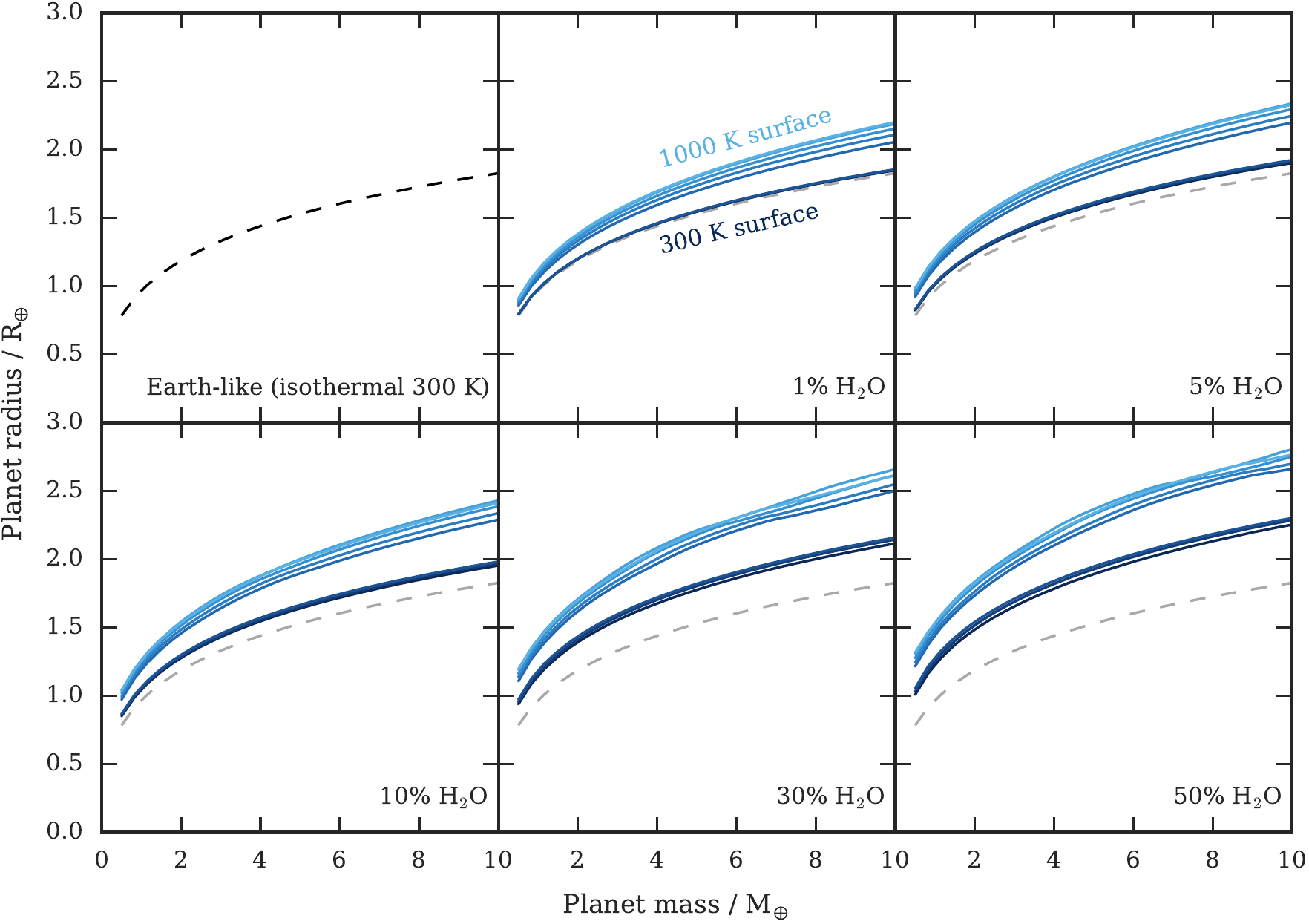}
\caption{Dependence of radii on water mass fraction. Even low-mass water
layers result in planets that are strongly affected by temperature
changes, especially when water on the surface is hot enough to be in the
vapour or supercritical phase. Here we show mass--radius relations for
multi-layer planets: an iron core with silicate and (in all but the
first panel) water layers. We show the Earth-like iron-silicate core in
each panel for comparison. All the watery planets are larger than the
dry case owing to the lower density of water. Surface temperature
variation affects the radius of a watery planet by a similar amount in
each case, and it can increase the radius by up to 25 per cent. Because
the iron and silicate layers are isothermal, this variation is due
solely to temperature effects in the water layer. We fixed the
silicate:iron mass ratio at 2:1 and set the surface pressure to
\(10^7\,\)Pa (100\(\,\)bar). The temperature contours are in steps of
100\(\,\)K.\label{fig:composition-variation}}
\end{figure*}

We find that changing the water content does not significantly affect
the temperature-dependent behaviour discussed in earlier sections (Fig.
\ref{fig:composition-variation}). We constructed planets with water,
silicate, and iron layers, fixing the silicate:iron mass ratio to the
Earth value of 2:1 and allowing the water shell to vary in mass. These
models correspond to an Earth-like nucleus with an extended water layer
at the surface.

The effects of surface temperature on radius are comparable in magnitude
across all our models with water layers, even when we set the water
layer mass to just 1 per cent of the mass of the entire planet. For a
10\(\,\)M\(_⊕\) super-Earth with a surface pressure of \(10^7\,\)Pa
(100\(\,\)bar), the radial change when the surface temperature increases
300 to 1000\(\,\)K is 0.5\(\,\)R\(_⊕\) (for a 50 per cent water planet)
and 0.4\(\,\)R\(_⊕\) (for a 1 per cent water planet). This similarity
holds across the entire range of planetary masses we considered. This
suggests that the bulk of the radius change comes from a water layer on
the surface whose density depends strongly on the surface temperature.

\subsection{Effect of temperature dependence on phase
structure}\label{phase-structure}

\begin{figure}
\centering
\includegraphics{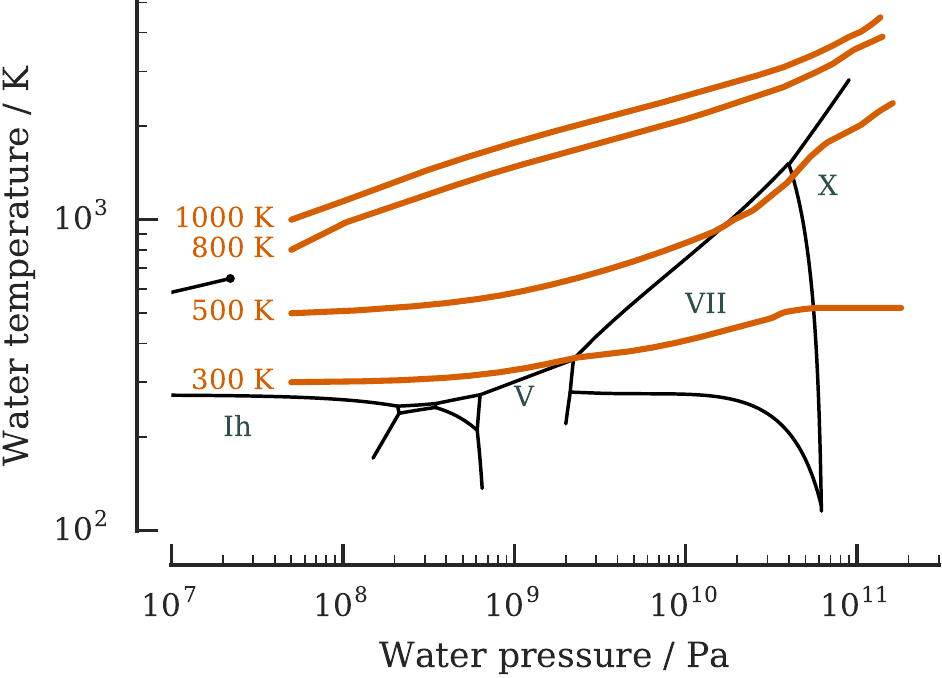}
\caption{Model pressure--temperature profiles. Increasing the surface
temperature means that more of the planet consists of superionic or
plasma phases of water, with the transition to high-pressure ice
happening deeper within the interior or not at all. Here we show
pressure--temperature profiles for 3\(\,\)M\(_⊕\) spheres of water with
a surface pressure of \(5×10^7\,\)Pa, which is beyond the critical
pressure. At a surface temperature of 300\(\,\)K, the planet consists of
liquid water over an ice VII core. But the ice VII phase may not be
present within the interior at higher surface temperatures. Instead, the
bulk of the planet consists of water in the superionic or plasma state.
This, combined with the low-density supercritical fluid at the surface,
results in an inflated radius.\label{fig:pressure-temperature-profiles}}
\end{figure}

Our adiabatic assumption provides for water layers which span different
phases depending on the pressure--temperature profile. As previously
noted, we observed a significant bifurcation in the mass--radius
diagrams when the surface temperature crossed the condensation curve of
water. As an example of how the surface temperature affects the
structure of a planet at pressures beyond the critical pressure of
water, Fig. \ref{fig:pressure-temperature-profiles} shows
pressure--temperature profiles for adiabatic spheres of water with
different surface temperatures. The planet contains a significant ice
VII component when the surface temperature is low. But at high surface
temperatures the centre of the planet may consist mostly of the
superionic or plasma phases of water, which are shown in Figs
\ref{fig:water-phases} and \ref{fig:eos-phase-space}.

Others have explored the layered phase structure of watery planets
\citep[e.g.][]{Zeng2014, Ehrenreich2006}. In particular,
\citet{Ehrenreich2006} included an analysis of radiogenic heating in
their models to assess the feasibility of having a liquid ocean under a
cold ice shell.

Though we have not assessed how the layered phase structure of a
planet's water layer might affect other properties of the planet, the
phase of water could be important for its potential to sustain
convective energy transport or magnetic fields \citep{Zeng2014}. We did
include a more complete treatment of the thermal expansion coefficient
\(α\) (calculating it directly from the equation of state) and the
variable heat capacity \(c_p\). This approach may result in a phase
structure that differs from other studies. In future we anticipate
investigating this more closely to assess whether these internal energy
sources are indeed sufficient to drive convection throughout our models:
is the assumption of a fully convective interior reasonable? As a first
indication of this, we consider the fact that we do not find major
deviations from the mass--radius relations of \citet{Valencia2007a}
(Fig. \ref{fig:valencia-mr-comparison}) to be promising. This is despite
the fact that they include conductive boundary layers in their models.

The phase structure is also of interest when we consider questions of
habitability. The properties of water change significantly near the
critical point: water becomes a low-dielectric fluid and a poor solvent
for polar substances \citep{Ansimov2004}. The nature of reactions
supported by water is also expected to change at high temperatures
\citep{Kruse2007}. We would expect significant changes in any kind of
life that could be found within these water layers, both when compared
with liquid oceans on Earth and when seen over time as the planet's
structure evolved.

\section{Conclusion and discussion}\label{conclusion-and-discussion}

In this paper we have presented planetary interior structure models of
water-rich super-Earths. The models incorporate a temperature-dependent
water equation of state and use an adiabatic treatment for the
temperature gradient. In doing so, we synthesized an updated equation of
state for water which attempts to capture all the relevant
temperature-dependent behaviour. We directly calculated the thermal
expansion coefficient \(α\) from the equation of state, rather than
treating it as a constant, and we used a variable heat capacity based on
experimental data. Our conclusions are as follows.

First, when one models a solid planet, adding a water layer comes with a
substantial thermal dependence. By this we mean that the temperature of
the planet may substantially alter the radius of the planet as the water
layer expands and contracts. Previous studies have shown that including
a temperature gradient in Earth-like planets produces a minimal change
in its radius \citep{Howe2014, Grasset2009, Seager2007}. We showed that
this assumption no longer holds once large water layers are considered,
even setting aside the unrealistic case of a 100 per cent water planet.
For example, consider the case of a 4\(\,\)M\(_⊕\) planet with an
Earth-like core underneath a water layer of 5 per cent of the planet's
total mass. If the surface pressure is \(10^7\,\)Pa (100\(\,\)bar), the
difference in the planet's radius when the surface is heated from
300\(\,\)K to 1000\(\,\)K is approximately 0.3\(\,\)R\(_⊕\) (Fig.
\ref{fig:composition-variation}). This effect is on top of any thermal
expansion of iron and silicate: our models treated the rocky layers as
isothermal. It is also in addition to any uncertainty in the equation of
state itself. Such changes in radii are significant considering that
current observations can already measure super-Earth radii to precisions
better than 0.1\(\,\)R\(_⊕\) (e.g.~Fig.
\ref{fig:isothermal-vs-adiabatic}).

The strength of the planet radius-temperature relation also depends on
the surface pressure. This is a result of the decreasing thermal
expansion of water with pressure: the coefficient of thermal expansion
is much smaller in high-pressure ice than in the liquid, vapour, or
supercritical fluid phases. At pressures of more than about
\(10^{10}\,\)Pa (\(100\,000\,\)bar) the radial temperature dependence
becomes irrelevant: the uncertainty in current planetary radius
measurements is larger than any conceivable radial change owing to
temperature effects, so more precise structural models may not be
useful. However, there is still a significant radial dependence on
temperature at lower surface pressures. At \(10^8\,\)Pa (1000\(\,\)bar),
a watery super-Earth with a surface temperature of 1000\(\,\)K can be up
to 0.1\(\,\)R\(_⊕\) larger than one with a surface temperature of
300\(\,\)K. It is therefore important to include temperature effects in
the interior models if an accurate radius is required as part of the
model.

This pressure dependence manifests itself most strongly below the
critical point of water. At pressures below this critical pressure, a
phase transition still exists between liquid and vapour. There is
therefore a bifurcation in the mass--radius diagram: a small increase in
surface temperature can causes a large change in radius (up to a factor
of two) as the surface water vaporises. We caution that it is likely not
appropriate to attempt to treat such vapour layers using our approach,
which is intended for interior structures. However, a lesser version of
this effect is still visible at higher pressures.

We consider the surface pressure as a free parameter in our models. In
principle, the surface pressure could be constrained through
spectroscopic observations of the planetary atmosphere, though such
observations are currently difficult for super-Earths. The surface
pressure is set by the depth beyond which atmospheric measurements can
no longer probe. \citet{Madhusudhan2015} discussed planets with
water-rich atmospheres, describing the use of measurements both in and
out of opacity windows to determine the atmospheric thickness. The
pressure to which these measurements probe varies from 0.1\(\,\)bar (in
regions of high opacity; that is, outside an atmospheric window) to
100\(\,\)bar (within such a window). Our models go beyond this pressure
range, and are therefore appropriate to treat the structure of the
planet below the observable opacity surface.

In the case of a volatile layer such as water, the line between interior
and atmosphere can become blurred. The picture is complicated by
atmospheric effects that can increase the opacity. If a cloud layer
forms in the atmosphere, the opacity surface may not necessarily be at
the same depth or pressure as any solid surface of the planet. Turbidity
effects around the critical point may also affect the opacity. It is for
this reason that high-temperature exoplanets are interesting: at higher
temperatures, a cloud deck is less likely to occur and atmospheric
measurements are therefore able to probe deeper. The
previously-mentioned opacity windows may therefore be able to provide a
view through the atmosphere to the planet's surface, or at least to a
point where the assumption of interior convective mixing is more likely
to hold.

We therefore conclude that, in some cases, planetary heating may alter
the interpretation of a planet's radius if a water layer is part of the
model. This is especially true if the planet consists entirely of water,
but this is an unlikely physical scenario. More importantly, the result
is still significant even if the surface of the water layer is at
moderately high pressures and lies underneath a heavy atmosphere. All
that is required for the water layer's density to change significantly
from the isothermal case is for a temperature increase of a few hundred
Kelvin. Moreover, even isothermal watery planets have some degree of
radial temperature dependence: up to 0.3\(\,\)R\(_⊕\) across the mass
range of super-Earths and in the temperature range of 300 to
1000\(\,\)K.

Understanding how the mass--radius relation can be affected by
temperature allows us to take the step of detecting and characterising
water-rich planets, taking their surface temperatures into account while
modelling them. This is an important precursor to narrow the search to
planets that would be considered more classically habitable. It will be
especially useful in the context of the next generation of super-Earths
expected to be found orbiting bright stars by missions such as PLATO
\citep{Rauer2014}, TESS \citep{Ricker2014} and CHEOPS \citep{Broeg2013}.
This approach is promising because it is linked to the characteristic
equilibrium temperature, which can be determined from observations of
the planet, and so can be included in analyses of populations of
planets. Through this we might better understand what proportion of
planets include substantial water content.

The temperature dependence is also important to take into account in
approaches such as that of \citet{Kipping2013}, where a watery interior
model is used to place a lower bound on the atmospheric height of an
observed planet. We have shown that the radius of an adiabatic watery
planet may be significantly higher than the zero-temperature or
isothermal case. Incorporating a surface temperature estimate into this
approach should therefore give better constraints.

From an observational perspective, these results are most interesting at
intermediate pressures. At low pressures (\(10^5\,\)Pa or 1\(\,\)bar) we
cannot claim that we accurately capture the behaviour of what is now
essentially an atmosphere, because we include no prescription for
radiative energy transport in our models. At high pressures
(\(10^{10}\,\)Pa or \(100\,000\,\)bar) any temperature dependence in the
water equation of state disappears. The physical scenario most relevant
for these models is therefore that of a water layer (ocean, ice or
supercritical fluid) underneath a thin or moderate atmosphere. Others
such as \citet{Rogers2010a} have already included volatile layers on top
of interior structure models. Adding more complete temperature
dependence to the interior portion of these planetary models is a
worthwhile future direction if we wish to treat them as water-rich.

We look forward to two developments in particular. The first is improved
atmospheric characterisation and modelling, which will provide useful
pressure and temperature boundary conditions at the base of the
atmosphere. The question of interior--atmospheric interactions is a rich
one that is only starting to be explored. Integrating atmospheric and
interior models promises progress on questions about surface chemistry,
outgassing and other processes that can shape the atmosphere of a
planet. The second development that will make use of this work is
improved spectral resolution of atmospheric observations, and in
particular the ability to seek out atmospheric windows
\citep{Madhusudhan2015}. By observing at wavelengths which pass through
the atmosphere, we can in principle directly measure the radius of any
solid interior underneath that atmosphere and thus have a better
starting point for interpreting the interior structure.

\section*{Acknowledgements}\label{acknowledgements}
\addcontentsline{toc}{section}{Acknowledgements}

We thank the anonymous reviewer for an insightful review and Christopher
Tout for helpful discussions and comments. ST gratefully acknowledges
support from the Royal Society of New Zealand.

\begin{center}\rule{0.5\linewidth}{\linethickness}\end{center}

\bibliographystyle{mnras}
\bibliography{library}

\bsp
\label{lastpage}
\end{document}